\documentclass[%
  reprint, amsmath,amssymb, prb, aps, superscriptaddress]{revtex4-2}

\usepackage{graphicx}
\usepackage{dcolumn}
\usepackage{bm}
\usepackage{hyperref}
\hypersetup{colorlinks=true, linkcolor=blue, citecolor=blue, urlcolor=blue}
\usepackage{xcolor}
\usepackage{orcidlink}\usepackage{multirow}

\begin{document}

\title{Systematic comparison of approximations and functionals in first-principles calculations of aluminum-based III-V ferroelectric nitrides}
\author{Alejandro Mercado Tejerina\,\orcidlink{0000-0002-6683-9496}}
\affiliation{Smart Ferroic Materials Center, Arkansas Materials Institute and Department of Physics, \href{https://ror.org/05jbt9m15}{University of Arkansas}, Fayetteville, AR 72701, USA.}
\author{Peng Chen\,\orcidlink{0000-0001-9934-3481}}
\affiliation{Smart Ferroic Materials Center, Arkansas Materials Institute and Department of Physics, \href{https://ror.org/05jbt9m15}{University of Arkansas}, Fayetteville, AR 72701, USA.}
\affiliation{Department of Physics, \href{https://ror.org/04rctme81}{Guangdong Technion -- Israel Institute of Technology}, Shantou, Guangdong Province, China.}
\author{Keisuke Yazawa\,\orcidlink{0000-0002-5304-1696}}
\affiliation{Department of Metallurgical and Materials Engineering, \href{https://ror.org/04raf6v53}{Colorado School of Mines}, Golden, Colorado 80401, USA.}
\affiliation{Materials Science Center, \href{https://ror.org/036266993}{National Laboratory of the Rockies}, Golden, Colorado 80401, USA.}
\affiliation{Smart Ferroic Materials Center, Arkansas Materials Institute and Department of Mechanical Engineering, \href{https://ror.org/05jbt9m15}{University of Arkansas}, Fayetteville, AR 72701, USA.}
\author{Andriy Zakutayev\,\orcidlink{0000-0002-3054-5525}}
\affiliation{Materials Science Center, \href{https://ror.org/036266993}{National Laboratory of the Rockies}, Golden, Colorado 80401, USA.}
\author{Laurent Bellaiche\,\orcidlink{0000-0002-6181-2024}}
\affiliation{Smart Ferroic Materials Center, Arkansas Materials Institute and Department of Physics, \href{https://ror.org/05jbt9m15}{University of Arkansas}, Fayetteville, AR 72701, USA.}
\affiliation{Department of Materials Science and Engineering, \href{https://ror.org/04mhzgx49}{Tel Aviv University}, Ramat Aviv, Tel Aviv 6997801, Israel.}
\author{Charles Paillard\,\orcidlink{0000-0002-3337-2295}}
\affiliation{Smart Ferroic Materials Center, Arkansas Materials Institute and Department of Physics, \href{https://ror.org/05jbt9m15}{University of Arkansas}, Fayetteville, AR 72701, USA.}
\email{paillard@uark.edu}
\date{\today}

\begin{abstract}
  We revisit first-principles predictions of structural, ferroelectric, and electronic properties in aluminum-based III–V nitride alloys, focusing on Al\textsubscript{1-x}Sc\textsubscript{x}N and Al\textsubscript{1-x}B\textsubscript{x}N. Using density functional theory within a unified 48-atom supercell framework, we systematically assess the role of chemical disorder and exchange–correlation approximations by comparing the virtual crystal approximation (VCA) and special quasirandom structures (SQS), as well as PBE, PBESol, SCAN, and SCAN+rVV10 functionals. We demonstrate that, even amongst the similar PBE and PBESol functionals, large quantitative and qualitative differences emerge. In particular, VCA or SQS PBESol modeling (a popular functional) underestimates the stability domain of the ferroelectric wurtzite phase in Al\textsubscript{1-x}Sc\textsubscript{x}N compared to SQS PBE or SQS SCAN. We demonstrate that the 5-fold coordinated hexagonal phase predicted in 2002 by Farrer and Bellaiche [\textit{Phys. Rev. B} 66, 201203] is a metastable state whose energy lies between that of the four-fold coordinated ferroelectric wurtzite phase and the energy of the six-fold coordinated rocksalt phase below the transition point. In contrast, Al\textsubscript{1-x}B\textsubscript{x}N shows a much faster destabilization of the wurtzite ferroelectric phase, with bond breaking which strongly distorts the wurtzite structure (with enhanced polarization) and eventually favors a zincblende phase and a threefold coordinated hexagonal layer phase.
  Our analysis highlights the critical importance of both local disorder and exchange–correlation treatment in the quantitative prediction of the functional properties of III–V nitride ferroelectrics. 
\end{abstract}

\maketitle

\section{Introduction}

The discovery of ferroelectricity in III-V nitride semiconductor thin films~\cite{Fichtner2019} has represented a major shift from conventional ferroelectric perovskite oxides. Their exceptionally large spontaneous polarization~\cite{Fichtner2019} and large thermal stability~\cite{Islam2021}, low-temperature processing and compatibility with back-end-of-line semiconductor processing~\cite{Kim2024} have attracted a lot of attention for applications in non-volatile memories~\cite{Hu2025}, piezoelectric devices, electro-optic modulators~\cite{Yang2024,Wang2025_eo}, RF electronics and high-electron mobility transistors~\cite{Wang2024}.

From a theoretical perspective, first-principles calculations based on density functional theory (DFT) have played a central role in elucidating the structural energetics and polarization mechanisms in III–V nitrides. Early studies established the relative stability of the wurtzite (WZ), rocksalt (RS), hexagonal (HX) and zincblende (ZB) phases in binary nitride compounds and highlighted the importance of coordination and bonding in determining phase transitions and electronic properties~\cite{Farrer2002,Limpijumnong2001}. Alloying strategies of WZ polar binary nitrides -- particularly with Sc or B~\cite{Pike2025,CalderonV2024} -- have been extensively explored to tune the structural landscape and destabilize the WZ phase, lower the coercive field and enable switching. Interestingly, many \textit{ab-initio} studies to date have employed the PBESol~\cite{Perdew2008} or PBE~\cite{Perdew1996} exchange-correlation approximation to explore the ferroelectric properties of prototypical nitride ferroelectrics (Al,Sc)N or (Al,B)N~\cite{Yazawa2022,Calderon2023,Chen2025,Arras2026}. However, there are few comprehensive studies that examine the effect of the exchange-correlation functional on the predicted structural and energetic properties of disordered ternary III-V nitride ferroelectrics. In particular, it is worth wondering if, as demonstrated in the case of classical ferroelectrics~\cite{Zhang2017}, metaGGAs such as SCAN may exhibit improved predictive accuracy in nitride ferroelectrics at a moderate computing cost.  Furthermore, most {\it ab-initio} studies have focused on using supercells (including SQS) to deal with the disordered nature of the metal distribution in AlN-based ferroelectrics. One may wonder if a more computationally friendly approach, such as the Virtual Crystal Approximation (VCA) would be justified, which would considerably shorten computation time and ease the construction of large-scale models~\cite{Chen2025}.


In this context, a systematic reassessment of first-principles methodologies is required to establish a consistent and predictive framework for nitride ferroelectrics. In this work, we revisit the structural, electronic, and ferroelectric properties of Al\textsubscript{1-x}Sc\textsubscript{x}N and Al\textsubscript{1-x}B\textsubscript{x}N using a unified computational approach. We explicitly compare VCA and SQS treatments of disorder and benchmark multiple exchange–correlation functionals, including PBE, PBESol, SCAN, and SCAN+rVV10. We find that SQS combined with SCAN provides the most consistent theoretical treatment of emerging nitride-based ferroelectric materials. 
Beyond methodological considerations, our results provide new insights into the fundamental physics governing these materials. We show that alloying with Sc and B leads to qualitatively different structural and ferroelectric responses, driven respectively by increased and decreased coordination tendencies. These findings establish clear structure–property relationships and offer guidance for the design of next-generation nitride ferroelectrics with tailored functionalities.

\section{Methods}

\begin{figure*}
    \centering
    \includegraphics{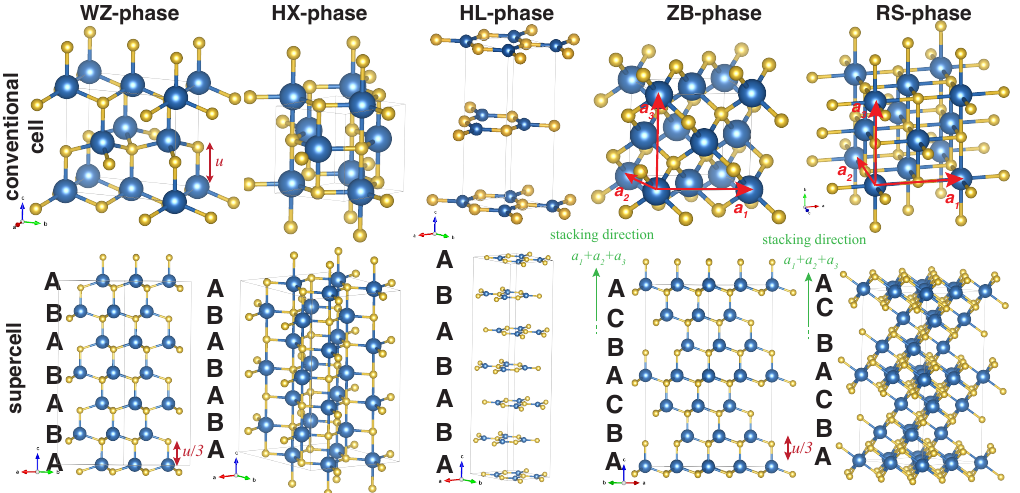}
    \caption{Sketch of the conventional cells (upper row) and common 48-atom supercell (lower row) of the wurtzite (WZ phase), hexagonal (HX phase), hexagonal-layered (HL phase), zincblende (ZB phase) and rocksalt phase (RS phase) considered in this work, visualized with the VESTA visualization software~\cite{Momma2008}. The blue spheres depict the metal atom (e.g. Al, Sc or B) and gold spheres represent the N atom. The notations A,B and C are used to describe the stacking of successive metallic atom planes (depicted in blue). Note that the AB stacking in the hexagonal systems (WZ, HX and HL) occurs in the $[001]$ direction, while the ABC stackings of the cubic phases refers to stacking along the $[111]$ direction in the conventional cubic cell.}
    \label{fig1:structures}
\end{figure*}

The structural complexity of III–V nitride alloys is amplified by the existence of multiple competing polymorphs with distinct coordination environments, ranging from three-fold (layered hexagonal) to six-fold (rocksalt). As highlighted in Figure~\ref{fig1:structures} of this work, these phases can be described within a unified supercell framework, enabling a direct comparison of their energetics and properties.
The polar wurtzite (WZ) phase and the zincblende (ZB) phases both show four-fold coordination of the metal atom. In contrast, III-V nitride semiconductors such as (Al,Sc)N or (Al,B)N can mainly favor five crystallographic phases: a polar wurtzite phase (denoted the WZ-phase), a layered hexagonal phase where the metal element is 3-fold coordinated (the HL structure; $h$-BN being a good representative of this phase), an hexagonal phase where the metal ion is five-fold coordinated (as predicted in ScN~\cite{Farrer2002} or MgO~\cite{Limpijumnong2001}, which we denote the HX-phase), a centrosymmetric cubic rocksalt structure (denoted the RS phase) and a zincblende cubic phase (hereafter labeled the ZB phase). Despite the hexagonal and cubic systems' unit cells being significantly different, one can find a common supercell able to describe all five phases on an equal footing. Such a supercell corresponds to a $2\times2\times3$ supercell of the wurtzite or hexagonal structure and thus contains 48-atoms. To obtain an equivalent supercell for the RS and ZB phases, one simply needs to transform the lattice vectors $(\bm{a}_1, \bm{a}_2, \bm{a}_3)$ of the cubic to the supercell vectors $(\bm{b}_1 = -\bm{a}_1 + \bm{a}_3, \bm{b}_2 = \bm{a}_1 - \bm{a}_2, \bm{b}_3 = 2\bm{a}_1 + 2\bm{a}_2 + 2\bm{a}_3)$. Figure~\ref{fig1:structures} depicts the supercells of all five phases, which highlights their key differences. Moreover, polar wurtzite WZ-phase of (Al,Sc)N or (Al,B)N differs from the zincblende ZB-phase by the stacking order: ABAB versus ABCABC. In both cases, the metal ion, in blue, is four-fold coordinated. The hexagonal and hexagonal-layered phases, despite presenting an ABAB stacking identical to the WZ-phase, have no out-of-plane or in-plane net dipole moment. Their metal ions are 5 and 3-fold coordinated, respectively. Finally, the rocksalt RS-phase possesses an ABCABC stacking, no dipoles, and its metal ions are sixfold-coordinated. Such vastly different available structures with big differences in coordination and thus chemical bonding are likely to present a rich array of very different electronic and optical properties. 

To address some of the questions related to the structural, electronic, electrical and functional properties of (Al,M)N (M=Sc or B), we use the 48-atom supercells depicted in Figure~\ref{fig1:structures} (unless otherwise stated) as the basis for our subsequent Density Functional Theory (DFT) calculations. DFT simulations performed here employ the \textsc{Vienna Ab Initio Software} (\textsc{Vasp}) package~\cite{Kresse1994,Kresse1996} with Projector Augmented Waves (PAW) pseudopotentials~\cite{Blochl1994,Kresse1999}. We set a plane wave cut-off of 620~eV and a $\Gamma$-centered $10\times10\times4$ sampling of the first Brillouin zone. In this work, we considered the PBE~\cite{Perdew1997}, PBESol~\cite{Perdew2008} and SCAN exchange-correlation functionals~\cite{Sun2015}. We also test for the effect of Van der Waals interactions using the rVV10 method~\cite{Sabatini2013} coupled with the SCAN exchange-correlation functional~\cite{Peng2016}.

To mimic the local chemical disorder, we create Special Quasi-random Structures~\cite{Zunger1990} using the 48-atom supercells depicted in Figure~\ref{fig1:structures} thanks to the \textsc{Alloy Theoretic Automated Toolkit}~\cite{vandeWalle2013}. Whenever specified, we also compared our results with the Virtual Crystal Approximation (VCA)~\cite{Ghosez2000}, as implemented in the \textsc{Abinit} plane-wave DFT code, using a plane wave cut-off of 60~Ha, and norm-conserved pseudopotentials. We use the VCA with PBE norm-conserved pseudopotentials implemented in \textsc{Abinit} because our earlier tests showed that the VCA method developed for PAW or ultra-soft pseudopotentials (as implemented in \textsc{Vasp})~\cite{Bellaiche2000} yields large remaining forces on the Sc and Al atoms, owing to the vastly different outer valence states of these species. In fact, special pseudopotentials for Al, including $2s$ and $2p$ electrons as valence were needed to achieve reasonable results (for instance, having an insulating ground state) in (Al,Sc)N with the VCA method.

\section{Energetics}

Let us start discussing the Kohn-Sham energy of the various phases in Al\textsubscript{1-x}M\textsubscript{x}N (M=Sc or B) for various exchange-correlation functionals and methods to mimic the local disorder (VCA versus SQS).

\begin{figure*}
    \centering
    \includegraphics[width=6.6in]{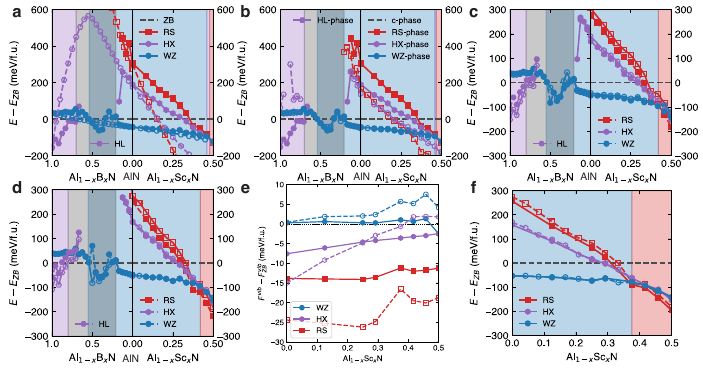}
    \caption{Energy of the WZ, RS, HX and HL phases with respect to the ZB phase, calculated using (a) PBE exchange correlation functional with VCA (open symbols) and SQS (plain symbols) treatment of chemical disorder, (b) SQS structures with PBESol (open symbols) and PBE (filled symbols), (c) SQS structures with PBE (open symbols) and SCAN (filled symbols), and (d) SQS structures with the SCAN (open symbols) and SCAN+rVV10 (filled symbols). In (e), we plot the vibrational free energy at 0~K (plain lines and filled symbols) and at 900~K (open symbols, dashed lines).}
    \label{fig2:phase_diagram}
\end{figure*}

\subsection{Virtual Crystal Approximation \textit{vs} Special Quasirandom Structures}

In Figure~\ref{fig2:phase_diagram}(a), we present the energy of the relaxed WZ, RS and HX phases of (Al,Sc)N in the right panel and the energy of the WZ, RS, HX and HL phases of (Al,B)N in the left panel, calculated with SQS supercells (plain symbols) and with VCA cells (open symbols) within the PBE exchange-correlation functional. All energies are reported with respect to the calculated ZB structure of (Al,Sc)N and (Al,B)N, respectively. 

\paragraph{Aluminum scandium nitride.} Let us first discuss the case of (Al,Sc)N (right panel in Figure~\ref{fig2:phase_diagram}(a)). We observe that the VCA and SQS calculations qualitatively give the same ground state results: at low concentration of scandium, the ferroelectric WZ phase (blue circles) is more stable than the ZB (dashed black line), the RS (red squares) and the HX (purple diamond) phases. Note that the HL phase is completely unstable in (Al,Sc)N and immediately relaxes in the HX phase. At large concentrations of scandium, in both cases, the non-polar RS phase becomes more energetically favorable. Note also that the WZ, HX and RS phases are nearly degenerate near the transition point, indicating the possible occurrence of a triple point in the phase diagram.
We note, however, some important quantitative differences with experimental observations. For instance, in the VCA case, we estimated (by linear interpolation) the crossing point between the energy of the WZ and RS phase to be around 19\% scandium content. This is a much smaller concentration than has been reported in experiments. For instance, Fichtner \textit{et al.} report square ferroelectric $P-E$ loops up to 43\% of scandium~\cite{Fichtner2019}; Yazawa \textit{et al.} report ferroelectricity in (Al,Sc)N films up to 35\% Sc content, beyond which the appearance of the RS phase, coexisting with WZ, destroys the ferroelectric properties~\cite{Yazawa2022}.

In stark contrast, the SQS method (filled symbols in Figure~\ref{fig2:phase_diagram}(a)) shows better \textit{quantitative} agreement with the experimental results reported in the literature. While we still observe a transition between the WZ (at low Sc concentration) and RS (at large Sc concentration), through linear interpolation of the WZ and RS energies near the transition region, we estimate that the critical concentration at which the RS phase becomes more energetically favorable is $x_c \approx 47.8$\% of scandium (compared to the critical concentration of 19.0\% in VCA, see Figure~\ref{fig2:phase_diagram}(a)). We checked that SQS results were independent of whether we use norm-conserved pseudopotentials in \textsc{Abinit} or PAW pseudopotentials in \textsc{Vasp} (see Supplemental Material~\cite{SM}). However, we note that the SQS structures with the PBE exchange-correlation functional also predict (from linear interpolation of the results) that the HX phase may become more stable than the WZ phase at a slightly lower concentration of 45.8\%. In fact, this result is correlated with the fact that an initial WZ structure with 45.8\% Sc content relaxes to a structure with a $c/a$ close to 1.2 (characteristic of the HX phase), rather than 1.6 (as expected in a WZ structure), as can be seen in Figure~\ref{fig5:wz_lattice_constants}. This result indicates that the HX phase may be the most stable phase in an extremely narrow concentration region, which, as far as we know, has never been reported experimentally. Finally, let us point out that, in the SQS method employing the PBE exchange-correlation functional, we observe that (1) as in the VCA, the three HX, WZ and RS phases are nearly degenerate near the transition point and (2) already at 45.8\% the initial WZ SQS structure relaxes into an HX structure, indicating that the WZ structure corresponds to either an extremely shallow minimum of energy with little to no energy barrier with the HX structure, or is downright unstable.

Note also that one may wonder whether the fact that we constrain our SQS structures to have a certain cell shape (see lower panels in Figure~\ref{fig1:structures}) has an effect and/or whether SQS size significantly alters the results. We compared these 48-atom SQS structures with constrained lattice vectors to 48-atom SQS structures whose lattice vectors were also an optimizing parameter in the SQS generation step performed with the \textsc{Atat} package. In that case, we found that the critical concentration at which the RS phase became energetically lower than the WZ phase shifted toward higher Sc content, for every exchange-correlation functional (see Table~\ref{tab:sqs_tests}). There, it can be seen that the critical concentration is shifted by a mere 3.4\% in PBESol, when it increases by almost 10\% in SCAN+rvv10. These results also highlight that the relative stability domains of WZ and RS are quite sensitive to local atomic order, which warrants a comprehensive future study of the effect of short-range ordering in (Al,Sc)N. In contrast, our tests on SQS sizes do not evidence strong SQS size effects (see Supplemental Material~\cite{SM}).

\begin{table}[]
    \centering
    \begin{tabular}{c  c  c }
    \hline
         Exchange-correlation & SQS type & $x_c(WZ\rightarrow RS)$\\
         \hline
         \hline
         \multirow{2}{*}{PBESol} & constrained & 0.290 \\
                                 & free        & 0.324 \\
        \multirow{2}{*}{PBE}     & constrained & 0.478 \\
                                 & free        & 0.558 \\          
        \multirow{2}{*}{SCAN}    & constrained & 0.433 \\
                                 & free        & 0.509 \\ 
        \multirow{2}{*}{SCAN+rvv10} & constrained & 0.359 \\
                                    & free        & 0.456 \\
        \hline
    \end{tabular}
    \caption{Interpolated critical concentration at which the energy of the WZ phase becomes larger than the RS phase in 48-atom SQS supercells with ("constrained") and without ("free") constrained shape.}
    \label{tab:sqs_tests}
\end{table}

\paragraph{Aluminum boron nitride.} Let us now turn our attention to the case of (Al,B)N and compare the energy of various phases predicted by VCA and SQS within the PBE approximation to the exchange-correlation functional. In the left side of Figure~\ref{fig2:phase_diagram}(a), one can see a markedly different picture compared to (Al,Sc)N. Let us first focus on the VCA data (open symbols in Figure~\ref{fig2:phase_diagram}(a)). We observe that the WZ phase has the lowest energy until 65.5\% boron content. Subsequently, the ZB phase (and not the RS phase as in (Al,Sc)N) becomes lower in energy. Finally, for $x > 94.4$\% a hexagonal layered phase (HL phase), characterized by a giant $c/a$ ratio exceeding 3, is the most energetically favored phase. Although the HL phase is routinely synthesized and observed at ambient temperature and pressure~\cite{Pease1950,Corrigan1975}, calculations beyond PBE have shown that the actual 0~K ground state should be a ZB phase~\cite{Cazorla2019}.

\begin{figure}
    \centering
    \includegraphics[width=3.3in]{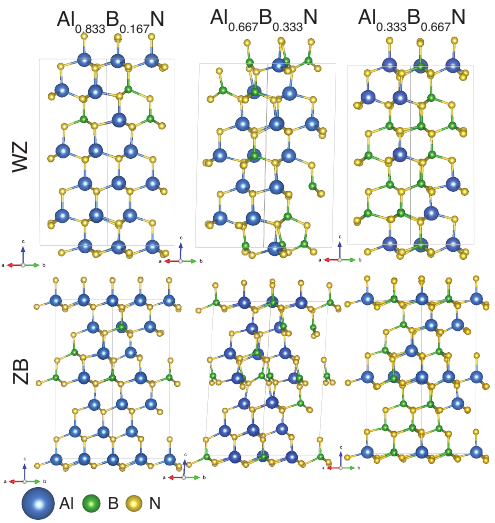}
    \caption{Relaxed SQS structures with the PBE exchange-correlation functional with the (top row) WZ and (bottom row) ZB initial structure, for (left) $x=0.167$, (middle) $x=0.333$ and (right) $x=0.667$ in Al\textsubscript{1-x}B\textsubscript{x}N.}
    \label{fig3:weird_phases_albn}
\end{figure}

Let us now describe what happens in (Al,B)N calculated within the SQS approximation (filled symbols in Figure~\ref{fig2:phase_diagram}(a)). We see, similar to the VCA, that the energy of the RS and HX phases strongly increases with respect to the WZ and ZB states as the B content increases. In fact, beyond 8-12.5\% boron concentration, it becomes difficult to even stabilize a structure remotely resembling the RS phase in the SQS, meanwhile the HX phase relaxes into the WZ phase beyond $x=12.5$\%. We subsequently do not discuss the RS and HX phases in (Al,B)N in the remainder of this article.
The most stable SQS phases (WZ, ZB and HL) bear qualitative similarities to the VCA in the low ($x < 20$\%) and high ($x > 50$\%) concentrations. The WZ phase has the lowest energy up to $x = 16.7$\%. Beyond $x = 54.2$\%, the ZB phase becomes lower in energy than all other considered phases. Finally, after $x=70.8$\%, the HL phases becomes the ground state. We note, however, that in the intermediate regime ($20 < x < 50$\%), the relative energy of the (initially constructed) WZ and ZB phases is very non-monotonous after structural relaxation. Examining the structures in more details revealed that this originates from bond breaking of some B atoms, which prefer to adopt a threefold coordinated environment (see middle panel in Figure~\ref{fig3:weird_phases_albn}). Surprisingly, at large B content, this effect is no longer observed and the WZ and ZB structures are recovered (see right panel in Figure~\ref{fig3:weird_phases_albn}). Note that, in this intermediate regime, the distorted WZ and ZB structures are dynamically stable, \textit{i.e.} no imaginary optical or acoustic phonons, calculated by finite differences in the 48-atom supercells are revealed; in contrast, regular WZ and ZB structures in (Al,B)N for $12.5 < x < 50$\% clearly exhibit unstable phonon modes (see, for instance, Figure S4 in the Supplemental Material~\cite{SM}) precisely associated with the boron bond breakings evidenced in the relaxed stable structures depicted in the middle panel of Figure~\ref{fig3:weird_phases_albn}.

Recent experimental results reveal that ferroelectricity is observed up to $x\approx 0.19$, while no polarization is detected in $P-E$ loops measurements at $x=0.20$~\cite{Hayden2021}. Meanwhile, a constant $c/a \approx 1.6$ is measured by X-ray diffraction up to $x=0.20$~\cite{Hayden2021}, which correlates well with our SQS and VCA calculations showing a regular and lowest energy WZ structure in this concentration range (see Figure~\ref{fig4:phase_diagram_summary}).

\subsection{Exchange-correlation functional}

The previous results, comparing VCA and SQS structures, clearly show that VCA severely underestimates the stability window of the ferroelectric WZ phase in (Al,Sc)N. In this subsection, we solely focus on SQS structures and explore different exchange-correlation functionals, namely PBE~\cite{Perdew1997}, PBESol~\cite{Perdew2008}, SCAN~\cite{Sun2015} and consider the effect of Van der Waals interactions at the rVV10~\cite{Sabatini2013,Peng2016} level of approximation.

\subsubsection{PBE versus PBESol}

In Figure~\ref{fig2:phase_diagram}(b), we depict the energy of the relevant phases in (Al,Sc)N (right panel) and (Al,B)N (left panel) calculated with the PBESol (open symbols) and PBE (plain symbols) exchange correlation. Quite surprisingly, it can be observed that PBESol and PBE give very different quantitative results in (Al,Sc)N. Specifically, the RS phase becomes lower in energy near $x_c \approx 29.1$\% Sc content using PBESol compared to $x_c \approx 47.8$\% with PBE. It thus seems that PBESol significantly underestimates the critical concentration from the WZ to the RS phase compared to experimental reports observing a WZ phase up to 43\% scandium content~\cite{Fichtner2019}. We also note, in the case of PBESol, that the energies of the RS and the HX phase are nearly degenerate (about 24.8~meV/f.u. in pure AlN), meanwhile the energy of the RS phase is originally significantly higher than the HX phase in the PBE approximation (115.5~meV/f.u. in AlN). We further observe that it is mostly the 6-fold coordinated RS phase whose energy seems poorly described in PBESol; in contrast, the relative energy of the 4-fold coordinated WZ and ZB phases is almost identical in PBESol and PBE, while the energy of the 5-fold coordinated HX phase with respect to the ZB structure is slightly higher in PBE (by 47~meV/f.u.).

Looking at Al\textsubscript{1-x}B\textsubscript{x}N in the left panel of Figure~\ref{fig2:phase_diagram}(b), we observe that PBE and PBESol lead to even some \textit{qualitative} differences at large boron concentration. Indeed, while PBE predicts that the 3-fold coordinated HL phase is the most stable above $x\approx 70.1$\%, PBESol indicates that the ZB phase is the lowest energy state from $x \approx 53.9$\% all the way to pure BN. We note that, as in the case of Al\textsubscript{1-x}Sc\textsubscript{x}N, the relative energy of the WZ and ZB phases is almost identical in PBEsol and PBE.

It may appear surprising that PBE and PBESol exchange-correlation functionals may give different energetic results. Yet, past studies have shown that, although PBESol gave more accurate lattice constants in solids, PBE tended to provide a better description of energy quantities such as the cohesive energy~\cite{Czonka2009}.

\subsubsection{PBE versus SCAN}

Let us now investigate how the semi-local GGA PBE exchange-correlation functional compares with the metaGGA SCAN functional. The latter was deemed almost as precise as the B1-WC hybrid functional over a wide range of ferroelectric materials, primarily oxides (e.g. BaTiO\textsubscript{3}, BiFeO\textsubscript{3}) and molecular ferroelectrics (e.g. KH\textsubscript{2}PO\textsubscript{4}, PhMDA)~\cite{Zhang2017}. Zhang \textit{et al.} further evidenced a systematic improvement of lattice constants, polar distortion and electronic polarizability over the Local Density Approximation (LDA) and PBE~\cite{Zhang2017}. Quite interestingly, PBE (empty symbols) and the SCAN meta-GGA (filled symbols) agree extremely well both qualitatively and quantitatively, as demonstrated in (Al,B)N and (Al,Sc)N in Figure~\ref{fig2:phase_diagram}(c). Only minor quantitative differences are observed. Namely, in Al\textsubscript{1-x}Sc\textsubscript{x}N, the RS phase becomes the lowest energy phase at $x_c = 0.434$ with the SCAN meta-GGA, while it is slightly higher in PBE ($x_c = 0.478$), see Figure~\ref{fig4:phase_diagram_summary}. In (Al,B)N, the stability window of the HL phase is slightly reduced (starting at $x\geq 0.808$ in SCAN versus $x\geq 0.701$ in PBE), meanwhile the ZB phase becomes lower in energy than the WZ phase at a similar concentration of boron (54.1\% versus 53.9\% for SCAN and PBE respectively).

\subsubsection{Effect of Van der Waals interactions}

Finally, we look at the role of potential van der Waals interactions in (Al,Sc)N and (Al,B)N. Although we do not expect to play a major role in (Al,Sc)N, the existence of the hexagonal layered phase in BN indicates that these effects may be significant in (Al,B)N. Here, van der Waals interactions are explored by combining the rVV10 method with the SCAN exchange-correlation functional in Figure~\ref{fig2:phase_diagram}(d). Looking at the case of (Al,Sc)N (right panel), it can be seen that the relative energy curves of the HX and WZ phases with respect to the ZB phase are virtually indistinguishable between the SCAN (open symbols) and SCAN+rVV10 cases. On the other hand, van der Waals interactions shift down the RS energy curve by about 25~meV/f.u. This results (see Figure~\ref{fig4:phase_diagram_summary}) in the expected stability of the RS phase occurring at a smaller concentration $x_c \approx 0.359$ of Sc in SCAN+rVV10 compared to SCAN ($x_c \approx 0.434$).

Similarly, in the case of (Al,B)N, the stability window of the WZ phase does not change significantly whether SCAN or SCAN+rVV10 is used (see left panel in Figure~\ref{fig2:phase_diagram}(d)). However, the energy of the HL phase is shifted upward by approximately 50~meV/f.u. when switching on the rVV10 correction of energies and forces. The van der Waals interactions therefore reduce the domain of stability of the HL phase in (Al,B)N, which is expected to be the lowest energy state in SCAN+rVV10 for $x\geq 0.878$ (versus $x\geq 0.808$ with the SCAN exchange-correlation only).

\subsection{Effect of lattice vibrations}

Due to the quasi-degeneracy of the WZ, HX and RS phases near the transition region, one may wonder whether lattice vibrations such as zero-point vibration energy may significantly alter the phase stability and critical concentrations. Using the \textsc{Phonopy} software package~\cite{phonopy-phono3py-JPSJ}, we calculated the lattice vibrational modes of the 48-atom supercells in (Al,Sc)N and, in the harmonic approximation, the contribution of phonons to the free energy, $F_{vib}$. Our results, shown in Figure~\ref{fig2:phase_diagram}(e), show that at 0~K, the contribution of vibrational modes to the free energy favor the RS phase over the WZ phase by about 15~meV/f.u. At a finite temperature of 900~K (dashed lines in Figure~\ref{fig2:phase_diagram}(e)), which is representative of a rather large growth temperature for (Al,Sc)N thin films~\cite{Talley2018}, the vibrational free energy of the RS phase is further lowered compared to that of the HX and WZ phases. Overall, lattice vibrational modes favor the RS phases and slightly shift the critical concentration at which the free energy of the RS phase becomes lower to $x_c\approx0.375$ (see Figure~\ref{fig2:phase_diagram}(f) and Figure~\ref{fig4:phase_diagram_summary}). We note, in Figure~\ref{fig2:phase_diagram}(f), that the RS phase maintains a quasi-degeneracy with the WZ between 37.5\% and 45.8\% Sc-content, which likely explains the large range of phase co-existence reported experimentally~\cite{Talley2018}.

Furthermore, we note that the lowest optical phonon modes in AlScN retain a real frequency (calculated using SCAN, see Supplemental Material Figure S4~\cite{SM}). Consequently, the HX, WZ, RS and ZB are all dynamically stable phases. Furthermore, the relaxation of the WZ phase into the HX phase at large Sc content is not driven by a soft optical mode, but rather by a significant lowering of the energy barrier between the HX and WZ local energy minima, consistent with recent phenomenological models~\cite{Chen2025}.



\subsection{Summary}

\begin{figure}
    \centering
    \includegraphics{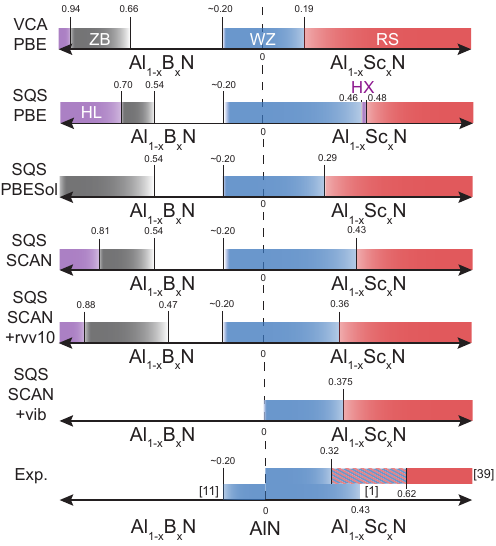}
    \caption{Most stable phase obtained from total Kohn-Sham energy calculations with different exchange-correlation functionals, alloying method (SQS or VCA). Vibrational contributions are included at the SCAN level of approximation and denoted SQS SCAN+vib. Experimental data taken from Refs~\cite{Fichtner2019, Talley2018, CalderonV2024}.}
    \label{fig4:phase_diagram_summary}
\end{figure}

Early experimental reports, by Fichtner et al. \cite{Fichtner2019}, indicate that the ferroelectric WZ phase has been observed up to 43\% scandium content in (Al,Sc)N films. As a result, it may indicate that the SCAN exchange-correlation functional applied on Special Quasirandom Structures may be more appropriate to model the ferroelectric properties of (Al,Sc)N. This would be consistent with earlier theoretical studies that demonstrated that SCAN performed better than GGA or LDA (and even as well as some hybrid functionals) in describing the structural and energetic properties of a wide array of ferroelectric materials~\cite{Zhang2017,Paul2017}. In contrast, PBESol applied to SQS structures or PBE applied to VCA structures predict a very low concentration transition from the WZ to the RS phase, which seems incompatible with Fichtner {\it et al.}'s report~ \cite{Fichtner2019} or the combinatorial experiments of Talley \textit{et al.}~\cite{Talley2018} which observe, depending on deposition conditions, a coexistence of the WZ and RS phases for concentrations of scandium ranging from 32 to 62\% (see Figure~\ref{fig4:phase_diagram_summary}). Note that most ferroelectric nitride films are grown using out-of-equilibrium methods such as sputtering, which explains why different studies may find different stability windows for the ferroelectric WZ and non-polar RS phases. 

Surprisingly, the semi-local GGA exchange functional PBE predicts that the HX phase (generally predicted as metastable in III-V nitrides like ScN~\cite{Farrer2002}) may be the most stable phase in a very narrow concentration window (see Figure~\ref{fig4:phase_diagram_summary}) which, as far as we know, has never been observed experimentally. Finally, we notice that van der Waals interaction may significantly reduce the transition point (36\% according to Figure~\ref{fig4:phase_diagram_summary}) at which RS may become more stable than WZ, which appear to be incompatible with the observations of Fichtner {\it et al.}; in addition, the use of van der Waals dispersion corrections such as the rVV10 method introduces a significant computational overhead. Overall, it is evident that Special Quasirandom Structures should be preferred over the Virtual Crystal Approximation to model the effect of alloying in (Al,Sc)N. Over all chosen functionals,PBE and SCAN predict the largest stability window for the WZ phase; however, PBE predicts that the HX phase could be more stable than WZ and RS which, as far as we know, has never been observed experimentally. 
Therefore, the SCAN functional appears to be a better choice among all combinations studied based on energetic considerations alone and will be used in the remainder of this study unless stated otherwise. Lattice vibrational effects slightly lower the critical concentration, but maintain a quasi degeneracy of the WZ and RS phases over a significant range of concentrations. Strain effects in real thin films may also contribute to the extension of the WZ stability window, although previous studies on AlN/ScN superlattices indicate that extremely large strains are required for such effects to occur (see Supplemental Figure S4 of Ref.~\cite{PhysRevB.110.054101}) and phenomenological modeling of WZ AlScN evidenced little strain dependence of ferroelectric properties~\cite{Yazawa2022b}. Recent experimental results have shown, on the other hand, that a Lu buffer layer could extend the stability domain of WZ (Al,Sc)N~\cite{Hirata2025}. These results call for additional investigations of interfacial effects (beyond epitaxial strain) on thin film growth.

In the case of (Al,B)N, comprehensive experimental reports of structural phases are rare. We are mostly aware of the work of Calderon \textit{et al.}~\cite{CalderonV2024} which indicates that a ferroelectric WZ phase is observed till $x\approx 0.2$ boron content. This is consistent with our results across all exchange-correlation functionals, which indicate that a true WZ phase is the most stable energy phase up to 0.167. After that, between $x=0.167$ and $x=0.208$, the regular WZ structure of Al\textsubscript{1-x}B\textsubscript{x}N becomes \textit{dynamically} unstable, with multiple (but not all) B atoms becoming three-fold coordinated (see lower middle panel in  Figure~\ref{fig3:weird_phases_albn}). It may indicate that no true solid solution of (Al,B)N may be formed between $x=0.208$ and an higher composition where a true ZB phase may appear. Most functionals investigated (PBE, PBESol and SCAN) indicate that the ZB structure becomes more stable than the (distorted or undistorted) WZ structure beyond $x = 0.54$. SCAN+rVV10 estimates a lower threshold ($x\approx 0.47$), and VCA PBE a larger one ($x \approx 0.66$), as depicted in Figure~\ref{fig4:phase_diagram_summary}. For rich boron concentrations, we note significant qualitative differences between exchange-correlation functionals. PBESol, for instance, predicts that the ZB phase is most stable for all concentrations $x\geq 0.54$. In contrast, VCA PBE or SQS PBE, SCAN and SCAN+rVV10 predict that the HL phase has lowest energy. All tested exchange-correlation functionals indicate that increasing the boron content in (Al,B)N promotes low coordination phases and strongly destabilizes phases with high coordination. This may relate to recent predictions of THz-driven bulk-to-layered transformations in BN~\cite{Liu2026}.

Finding HL structure as the ground state seems to contradict thermodynamical equilibrium phase diagrams that identify the ZB phase as the most stable phase in BN~\cite{Solozhenko1999}, despite the fact that the HL phase is most commonly found in nature~\cite{Pease1950}. Cazorla \textit{et al.} theoretically showed that many-body methods or a method using the adiabatic-connection fluctuation-dissipation theorem in the random phase approximation were able to correctly describe ZB as the ground state of BN, owing to its reduced dimensionality (essentially 2D layered material)~\cite{Cazorla2019}. 
These methods are computationally intensive and, in practice, computationally intractable for our 48-atom SQS. As a result, owing to our interest in the ferroelectric properties (Al,B)N and by comparison with (Al,Sc)N, we chose the SCAN+rVV10 exchange-correlation functional to describe Al\textsubscript{1-x}B\textsubscript{x}N in the following (unless otherwise stated).

\section{Structural properties}
\label{sec:structure}

We now turn our attention to structural features of (Al,B)N and (Al,Sc)N. 

\subsection{Lattice constants}

\subsubsection{Ferroelectric wurtzite phase and exchange-correlation functionals}

\begin{figure}
    \centering
    \includegraphics[width=\linewidth]{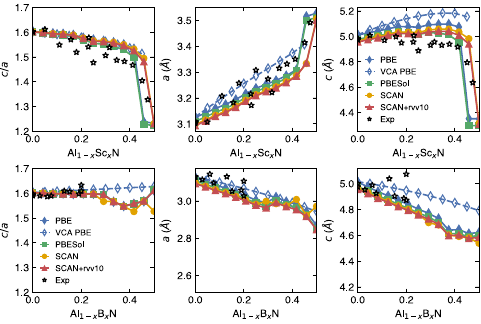}
    \caption{(left) $c/a$ ratio, (middle) $a$ lattice constant and (right) $c$ lattice constant in the WZ phase of (top) (Al,Sc)N and (bottom) (Al,B)N.}
    \label{fig5:wz_lattice_constants}
\end{figure}

\paragraph{Pure AlN.} Focusing on the ferroelectric WZ phase, Figure~\ref{fig5:wz_lattice_constants} shows its lattice constants along with the $c/a$ ratio in (Al,Sc)N and (Al,B)N. As expected, the WZ phase of AlN shows a $c/a$ in the range of 1.600 - 1.603 depending on the chosen exchange-correlation functional, which agrees almost perfectly with the measured values between 1.601-1.603~\cite{Satoh2022,Schulz1977,Akiyama2009} and is close to the ideal WZ $c/a$ ratio of $\sqrt{8/3} \approx 1.633$. Similarly, the $a$ and $c$ lattice constants (in particular those calculated with SCAN) are in very good quantitative agreement with experimentally reported values.

\paragraph{Wurtzite (Al,Sc)N.} In the case of WZ (Al,Sc)N (top row in Figure~\ref{fig5:wz_lattice_constants}), we observe a steady decrease of the $c/a$ ratio with Sc composition from 1.60 at $x=0$ to about 1.50 (PBESol) or 1.53 (PBE, SCAN, SCAN+rVV10), qualitatively consistent with experimental observations. A decrease in $c/a$ is generally associated with a decrease in spontaneous electrical polarization (as section~\ref{sec:polarization} evidences), which has recently been attributed to strong electromechanical atomistic couplings~\cite{Chen2025}. The top left panel in Figure~\ref{fig5:wz_lattice_constants} shows that, in general, every exchange correlation functional and method (SQS or VCA) tends to overestimate the $c/a$ ratio at large Sc concentrations (and thus, as we shall see in Section~\ref{sec:polarization}, overestimates the spontaneous polarization) with respect to experimental values~\cite{Satoh2022,Alvarez2023,Akiyama2009}. This quantitative discrepancy may be due to several factors. Firstly, our DFT calculations are performed at 0~K, while experimental observations reported here are conducted at room temperature. The electrical polarization and thus the $c/a$ ratio are expected to decrease with increasing temperature~\cite{Chen2025}, although both (Al,Sc)N and (Al,B)N seem to indicate a remarkable stability of their ferroelectric polarization up to elevated temperatures~\cite{Hayden2021,Islam2021}. Secondly, most available experimental data are reported on polycrystalline films grown by sputtering, while our DFT data are collected on perfect bulk monodomain monocrystals. It can be expected that defects and/or the interface with the substrate may reduce the electrical polarization and thus the $c/a$ ratio. We note, in the middle and right panels of Figure~\ref{fig5:wz_lattice_constants}, that the VCA calculated data overestimate both $a$ and $c$ in the WZ phase of (Al,Sc)N; in contrast, SQS structures appear to give a quantitatively reasonable description of $a$ and a moderate overestimation of $c$ in most of the concentration range explored here.

\paragraph{Wurtzite (Al,B)N.} WZ (Al,B)N presents an entirely different structural behavior than WZ (Al,Sc)N. As can be observed in the left lower panel in Figure~\ref{fig5:wz_lattice_constants}, most of the reported experimental values indicate a steady $c/a \approx 1.58 - 1.61$ up to 20\% boron concentration~\cite{Hayden2021,CalderonV2024}, which is quantitatively well reproduced by all exchange-correlation functionals and by both SQS and VCA. Surprisingly, VCA quantitatively \textit{and} qualitatively differs from SQS predictions of the individual lattice constants $a$ and $c$ (see middle and right lower panels in Figure~\ref{fig5:wz_lattice_constants}) although both exhibit similar $c/a$ ratio. In fact, VCA calculations grossly overestimate the $a$ and $c$ lattice constants in (Al,B)N compared to SQS, despite showing good relative agreement in describing the energy of the WZ phase in Figure~\ref{fig2:phase_diagram}(a).

We also observe, in Figure~\ref{fig5:wz_lattice_constants}, that (Al,B)N behaves qualitatively differently than (Al,Sc)N. In (Al,B)N, both the $a$ and $c$ lattice constants decrease with increasing boron concentration, likely due to the smaller covalent radius of boron (84~pm {\it versus} 121~pm for Al~\cite{Cordero2008}). In contrast, the $a$ lattice constant of (Al,Sc)N increases with increasing scandium concentration, while the $c$ lattice constant remains almost constant before suddenly dropping when the WZ relaxes in the HX phase beyond 45.8\% of B concentration.


\subsubsection{Comparison with other phases.}

As we intend to compare the WZ, RS, HX, HL and ZB phases on an equal footing, we report the lattice constant $a$ and $c$ of the equivalent WZ unit cell. This corresponds to reporting a third of the long axis of the supercell sketched in Figure~\ref{fig1:structures} for the $c$ lattice constant and half of the average length of the in-plane hexagonal axes for $a$. A consequence of this choice is that the cubic RS and ZB structures do not have a $c/a$ of 1, but rather close to $\sqrt{3}\approx 1.633$.

\begin{figure}
    \centering
    \includegraphics{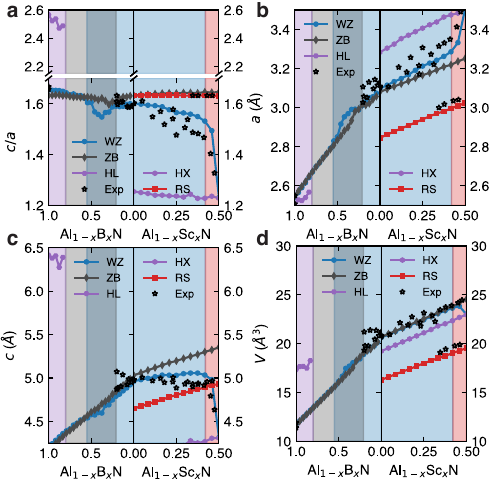}
    \caption{(a) $c/a$ ratio, (b) $a$ and (c) $c$ lattice constants, and (d) volume per formula unit in (Al,B)N (left panels) and (Al,Sc)N (right panel). Values are reported for the SCAN+rVV10 functional in (Al,B)N and the SCAN functional in (Al,Sc)N. Experimental values are taken from room temperature observations from Refs.~\cite{Akiyama2009,Schulz1977,Alvarez2023,Satoh2022,Yoshiasa2003,Hayden2021,CalderonV2024}.}
    \label{fig6:lattice_constants}
\end{figure}

In Figure~\ref{fig6:lattice_constants}(a-d), we present the $c/a$ ratio, lattice constants and volumes of the WZ, RS, HX, HL and ZB structures of (Al,Sc)N calculated with the SCAN exchange-correlation functional (right panels) and (Al,B)N calculated with the SCAN+rVV10 (left panels), as well as reported experimental values at room temperature~\cite{Akiyama2009,Schulz1977,Alvarez2023,Satoh2022,Yoshiasa2003,Hayden2021,CalderonV2024}. Let us first point out that the WZ and ZB phases (the latter emerging as a stable phase at large boron content) have similar lattice constants and almost identical volumes, either in Al\textsubscript{1-x}B\textsubscript{x}N or Al\textsubscript{1-x}Sc\textsubscript{x}N. This is likely the result of their similar bonding structure, made of a metal ion enclosed in a nitrogen tetrahedral cage (see Figure~\ref{fig6:lattice_constants}). 
Interestingly, we note that the highly coordinated phases (i.e., 5-fold coordinated HX phase and 6-fold coordinated RS phase) have the lowest volume at a given concentration in (Al,Sc)N. In contrast, the lower, 3-fold coordinated HL phase has a much higher volume than the 4-fold coordinated ZB and HX phases in (Al,B)N. 
Correlating these observations with the energy landscape revealed in Figure~\ref{fig2:phase_diagram} indicates that, as the average covalent radius of the metal ion increases, the volume increases and the emergence of higher coordination phases is favored (which is demonstrated by mixing Al and Sc); reciprocally, decreasing the covalent radius by mixing Al and B decreases the volume and favors phases with lower coordination of the metal ion. 



\subsection{Bond distribution}

\begin{figure}
    \centering
    \includegraphics[width=3.3in]{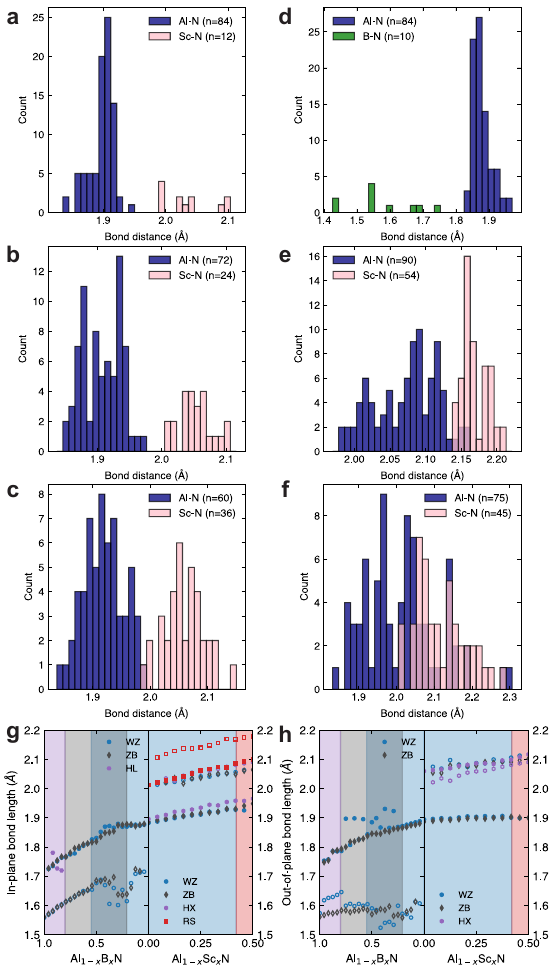}
    \caption{(a,b,c) Histograms of Al--N (navy) and Sc--N (pink) bond lengths in WZ Al\textsubscript{1-x}Sc\textsubscript{x}N, calculated with the SCAN functional for concentrations $x=0.125,0.250$ and $0.375$. (d) Histograms of Al--N (navy) and B--N (green) bond lengths in WZ Al\textsubscript{0.875}B\textsubscript{0.125}N, calculated with the SCAN+rVV10 functional. (e,f) Histograms of Al--N (navy) and Sc--N (pink) bond lengths in (e) RS and (f) HX Al\textsubscript{0.725}Sc\textsubscript{0.375}N. (g,h) Average (g) in-plane and (h) out-of-plane Al--N (filled symbols) and Sc--N or B--N (open symbols) bond lengths in (Al,B)N and (Al,Sc)N calculated using the SCAN+rVV10 and SCAN exchange-correlation functionals, respectively.}
    \label{fig7:bond_alscn}
\end{figure}

We calculated the length of Al-N, Sc-N and B-N bonds within a 2.5~\AA~ radius of the metal ion (Al, B or Sc) in Al\textsubscript{1-x}Sc\textsubscript{x}N and Al\textsubscript{1-x}B\textsubscript{x}N.
Because Sc is a transition metal with a significantly larger atomic radius than Al, the Sc--N average bond length in Al\textsubscript{1-x}Sc\textsubscript{x}N is about 0.2~\AA~ longer than the Al--N bonds across all four structures, as shown in Figure~\ref{fig7:bond_alscn}(a-c,e-h) \cite{Zhang2024,Bhattarai2024}. In contrast, Figure~\ref{fig7:bond_alscn}g shows that B--N bonds are, on average, 0.1~\AA~ shorter than Al--N bonds perpendicular to the $c$-axis, while they are 0.2-0.3~\AA~ shorter in the $c$-axis direction (see Figure~\ref{fig7:bond_alscn}h), reflecting the strong tendency of B to adopt a three-fold planar coordination.

Taking a deeper look at the distribution of bond lengths in (Al,Sc)N shows some remarkable features. In WZ (Al,Sc)N, Figure~\ref{fig7:bond_alscn}(a-c) reveals that the bond distribution of Al--N and Sc--N are well separated, except for a minor overlap at larger concentrations $x\geq 0.375$. Interestingly, the RS phase reveals a relatively large dispersion of Al--N bonds at all concentrations (see, for instance, Figure~\ref{fig7:bond_alscn}e for $x=0.375$); meanwhile, the Sc--N bond distribution is much narrower. Note  that Sc--N and Al--N bond length distributions start to overlap for $x\geq 0.208$ (not shown here). The HX phase, as depicted in Al\textsubscript{0.725}Sc\textsubscript{0.375}N in Figure~\ref{fig7:bond_alscn}(f), shows a marked difference from all other phases. Firstly, there is a significant overlap of the Sc--N and Al--N bond length distributions. Secondly, both the smallest and largest bonds tend to be Al--N bonds as soon as the concentration in Sc exceeds $x\geq 0.208$. Finally, the average bond length of Al--N and Sc--N are, contrary to the WZ, ZB and RS phases, almost the same in the HX phase. The HX phase is, in fact, the only phase where Al--N bonds pointing in the $c-$axis direction are, on average, longer than Sc--N bonds.

\subsection{Internal parameter $u$ in the WZ and ZB phases}

\begin{figure}
    \centering
    \includegraphics[width=3.3in]{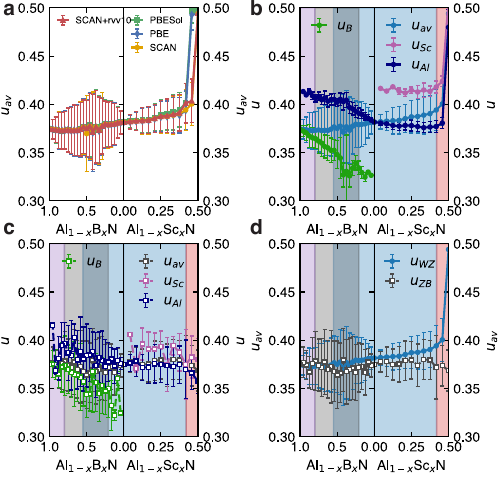}
    \caption{(a) Average internal parameter $u_{av}$ in WZ (Al,B)N (left panel) and WZ (Al,Sc)N (right panel) for different exchange-correlation functionals in SQS structures. (b) Internal parameter $u$ in WZ (Al,B)N calculated with SCAN+rVV10 and (Al,Sc)N calculated with SCAN, decomposed on metallic species in SQSs. (c) Internal parameter $u$ in ZB (Al,B)N and (Al,Sc)N, decomposed on metallic species in SQSs. (d) Comparison of the average internal parameter $u$ in the WZ and ZB SQSs of (Al,B)N and (Al,Sc)N. The error bars correspond to one standard deviation of the calculated internal parameters, and is an indication of the spread of internal parameters in the 48-atom SQS supercells.}
    \label{fig8:u}
\end{figure}

The WZ phase polar distortion is typically characterized by an internal parameter $u$ (see Figure~\ref{fig1:structures}), whose value is $0.5$ in the non-polar HX phase and is $3/8 = 0.375$ in the ideal WZ phase. We quantify, in the SQS supercells, the average internal parameter $u_{av}$ in the WZ phase by comparing the change in reduced atomic coordinates of M-N (M=Al, Sc or B) bonds in the polar direction with respect to their values in a hypothetical HX phase having the WZ cell lattice parameters (see Figure~\ref{fig1:structures}, lower left panel). The internal parameter $u_{av}$ has a close link to the electrical polarization, and has been used to build effective models~\cite{Chen2025}. We plot, in Figure~\ref{fig8:u}(a), the change in the average internal parameter calculated for different exchange correlation functionals in the WZ phase of Al\textsubscript{1-x}B\textsubscript{x}N and Al\textsubscript{1-x}Sc\textsubscript{x}N. One immediately notices opposite qualitative behaviors between Al\textsubscript{1-x}B\textsubscript{x}N and Al\textsubscript{1-x}Sc\textsubscript{x}N. The latter's internal parameter steadily increases, from $u_{av}\approx 0.382$ in pure AlN to 0.390 in Al\textsubscript{0.625}Sc\textsubscript{0.375}N, which points toward a slight decrease in spontaneous polarization (see section~\ref{sec:polarization}). Beyond $x\geq 0.375$, the average internal parameter starts to increase more rapidly to reach $u=0.5$, indicating that the polar distortion of the WZ structure is destabilized in favor of the metastable, non-polar HX phase. In contrast, $u_{av}$ decreases away from 0.5 as the concentration of boron increases in (Al,B)N. Starting from $u_{av} \approx 0.382$ in AlN, the average internal parameter essentially slightly decreases till $x=0.5$, where it stabilizes around the ideal WZ value of $0.375$. We note that there is little change in the structural features obtained under different exchange-correlation functionals, save for the earlier jump of $u_{av}$ from about 0.39 to 0.5 in (Al,B)N with the PBE functional. This correlates with our earlier energetic analysis, which showed that the HX phase became more stable than the WZ and RS phases in a narrow composition region (see Figure~\ref{fig4:phase_diagram_summary}).

To better understand the difference in qualitative behavior of (Al,B)N and (Al,Sc)N, we calculated the average internal parameter $u_M$ (M=Al, Sc or B) for each metallic species at each concentration in the WZ SQS supercells. Quite interestingly, the internal parameters for Al and Sc are approximately constant (approximately 0.380-0.375 and 0.413-0.418, respectively) for each concentration in (Al,Sc)N (see right panel of Figure~\ref{fig8:u}(b)), except when a transition to a HX phase is triggered at $x\approx 0.5$. As a result, the average internal parameter $u^{AlScN}_{av}(x) \approx (1-x)\bar{u}_{Al} + x \bar{u}_{Sc}$. 
In contrast, in (Al,B)N, $u_{Al}$ is initially linearly increasing from 0.382 in AlN to 0.404 in Al\textsubscript{0.583}B\textsubscript{0.417}N, subsequently followed by a much lower increase from 0.404 to 0.413 in WZ BN. $u_B$ follows a complementary trend: it is approximately constant or fluctuating between 0.324 and 0.330 until $x=0.417$, and subsequently increases up to 0.375 in WZ BN. The resulting average internal parameter, $u_{av}^{AlBN}$, very slightly decreases with increasing boron content.

Note that we can also devise an internal parameter for the ZB structure (see Figure~\ref{fig1:structures}) as it is structurally similar to the WZ structure, except that its MN\textsubscript{4} tetrahedra are stacked in an ABC fashion rather than ABAB. Looking at this internal parameter in the ZB phase in Figure~\ref{fig8:u}(c), we observe qualitatively similar trends to the WZ structure. $u_{Al}$ and $u_{Sc}$ are quasi constant with Sc concentration in (Al,Sc)N, meanwhile $u_{Al}$ and $u_B$ are both linearly increasing with boron content in (Al,B)N. Superimposing the average internal parameter of the ZB structure with that of the WZ structure in Figure~\ref{fig8:u}(d) reveals that the ZB structures have comparable but slightly lower average internal parameters. 



\section{Ferroelectric and dielectric properties}
\label{sec:polarization}

Structural parameters in section~\ref{sec:structure} already allude to qualitative trends in the polar state of (Al,B)N and (Al,Sc)N. In particular, we note that $c/a$ steadily decreases with increasing Sc content in WZ (meanwhile the internal parameter $u$ increases), suggesting qualitatively a \textit{decrease} in polarization. In contrast, (Al,B)N depicts an opposite trend, that is a slight increase of $c/a$ and an almost constant internal parameter $u$, which may indicate a stable or increasing spontaneous polarization.

\begin{figure}
    \centering
    \includegraphics[width=3.3in]{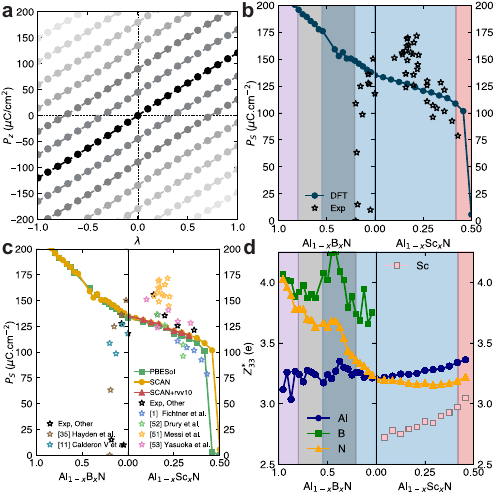}
    \caption{(a) Berry phase calculation of the electrical polarization versus the linear distortion $\lambda$ in the atomic positions between the WZ structure ($\lambda=\pm1)$ and a HX-like ($\lambda=0$) structure. Different shades of gray correspond to different quantum of polarization branches. (b) calculated spontaneous polarization in (left) (Al,B)N and (right) (Al,Sc)N, and (c) calculated using different exchange-correlation functionals. (d) Born charges in the polar direction using (left) SCAN+rVV10 in (Al,B)N and (right) SCAN in (Al,Sc)N.}
    \label{fig9:polarization}
\end{figure}

\subsection{Spontaneous polarization}

To obtain a more quantitative picture of the polar properties of (Al,Sc)N and (Al,B)N, we calculated the electrical polarization in the WZ phase of (Al,M)N (M=Sc or B) using the Berry phase approach~\cite{Resta1992,Vanderbilt1993,Resta1994}. Practically, and for each concentration of scandium or boron, we computed the polarization of structures whose ionic positions are scaled by increments of 10\% between those of a HX-like centrosymmetric phase and those of the polar, fully relaxed WZ structure. We maintain the lattice constant of the WZ phase while deforming the atomic positions in the SQSs. We use a HX-like centrosymmetric structure as a reference non-polar phase rather than the ZB structure~\cite{Dreyer2016}. We show, in Figure~\ref{fig9:polarization}(a), the polarization calculated for different values of distortions between the WZ ($\lambda = \pm 1$) and HX-like atomic positions ($\lambda = 0$) for Al\textsubscript{0.75}Sc\textsubscript{0.25}N, as well as the multiple branches of polarization in our 48-atom SQS supercell. Following the change of polarization on one branch between $\lambda=-1$ and $\lambda=+1$, we determine the spontaneous polarization for each concentration, which we plot in Figure~\ref{fig9:polarization}(b).

In Al\textsubscript{1-x}Sc\textsubscript{x}N, the polarization steadily decreases from 135.1~$\mu$C/cm\textsuperscript{2} in pure AlN to 108.7~$\mu$C/cm\textsuperscript{2} at $x=0.417$, \textit{i.e.} close to the boundary where the RS phase becomes lower in energy than the WZ, HX and ZB phases. We compare our results with experimental values reported in the literature~\cite{Fichtner2019,Messi2025,Drury2022,Yasuoka2020} using black stars in Figure~\ref{fig9:polarization}. One can observe a large spread in experimental values (see Figure~\ref{fig9:polarization}(b\&c). For instance, (Al,Sc)N films grown by Metal-Ion Synchronized High Power Impulse Magnetron Sputtering films~\cite{Messi2025} have reported giant polarization values in excess of 170~$\mu$C/cm\textsuperscript{2}; similarly, Molecular Beam Epitaxy epitaxially-grown films show large polarization values, above those predicted from Density Functional Theory (irrespective of the exchange-correlation functional, see Figure~\ref{fig9:polarization}(c)). A very recent preprint even reports a spontaneous polarization nearing 280~$\mu$C/cm\textsuperscript{2} in ultrathin 10~nm thick Al\textsubscript{0.68}Sc\textsubscript{0.32}N films grown by reactive pulsed DC magnetron
co-sputtering~\cite{Song2025}.
In contrast, earlier experimental reports from Fichtner \textit{et al.}~\cite{Fichtner2019} and Drury~\textit{et al.}~\cite{Drury2022} of films grown by DC and radiofrequency reactive magnetron sputtering have, overall, close or lower polarization values than predicted in our calculations. Short of the transition point from the WZ structure to the HX or RS phase, all exchange-correlation functionals yield similar quantitative results (see Figure~\ref{fig9:polarization}(c)). Potential origin of the discrepancy are (1) the thickness of the films and/or potential interfacial and depolarizing effects, (2) epitaxial strain, which earlier {\it ab-initio} prediction recognized as a strong tuning parameter of the spontaneous polarization in AlN/ScN superlattices~\cite{Jiang2019}, (3) large concentrations of defects or (4) incomplete saturation of P-E experimental loops due to the very large coercive fields. Note that our calculations agree well with past computational reports~\cite{Lee2025algdn}. 

Interestingly, our {\it ab-initio} calculations in Al\textsubscript{1-x}B\textsubscript{x}N indicate, for every exchange-correlation functional employed in this work, that the spontaneous polarization of the WZ phase strongly increases with boron concentration (see Figures~\ref{fig9:polarization}(b\& c)). Interestingly, some experimental reports seems to correlate the fact that, at low concentration of boron (of the order of a few percent), the spontaneous polarization does increase~\cite{CalderonV2024,Hayden2021}. We note that Hayden \textit{et al.} measured no polarization in Al\textsubscript{0.8}B\textsubscript{0.2}N films, which corresponds to the concentration where strong bond reconfiguration occurs and distorts the WZ and ZB structures according to our calculations. Note also that other experiments, such as MBE grown thin films, report much lower polarization values (10-15~$\mu$C/cm\textsuperscript{2}) even at both low (4.7\%) and intermediate concentrations of boron (18\%)~\cite{Savant2024}, which may be the result from strong wake up effects evidenced in the literature~\cite{Zhu2022}.

\subsection{Born charges}

We also calculated the Born effective charge in a structure having the WZ lattice constants, but with atoms in centrosymmetric positions (i.e., effectively, the reduced coordinates of the atoms correspond to that of the HX phase). The Born effective charges quantify the contribution of particular ions to the overall polarization. We present, in Figure~\ref{fig9:polarization}d, the average Born charges $Z_{33}^{*}$ along the polar direction for (Al,B)N and (Al,Sc)N calculated using the SCAN+rVV10 and SCAN functional, respectively. The results show drastically different qualitative behavior in (Al,Sc)N and (Al,B)N.

In (Al,Sc)N, the Born effective charges of both Al and Sc increase with concentration from 3.21 and 2.70 at $x=0.042$ to 3.36 and 3.04 at $x=0.458$, respectively. Accordingly, the Born effective charge of nitrogen remains approximately constant (in the range of $-3.15$ to $-3.20$) over the range of concentrations explored here.

In contrast, the magnitude of the Born effective charge of nitrogen strongly increases in (Al,B)N as the boron content increases. This is mainly driven by the large Born effective charge of B, whose value increases from 3.75 to beyond 4 in pure BN. In contrast, the Born effective charge of Al remains comparatively unchanged (with a slight decrease from 3.20 to 3.11). We note some strong changes of the Born charge in the region $0.208 \leq x \leq 0.592$, likely due to strong bond reconfiguration as discussed earlier and depicted in Figure~\ref{fig3:weird_phases_albn}.

Note that the polarization estimated by Born charges in the HX-like reference phase leads to an overestimation of the electrical polarization by about 10\% in (Al,Sc)N (see Figure S5 in the Supplemental Material~\cite{SM}) and even larger in (Al,B)N at high concentration of boron. This indicates that Born charges are likely not constant along the HX-like to WZ distortion path.











\section{Conclusion}

In this work, we have performed a comprehensive first-principles investigation of the structural, ferroelectric, and electronic properties of aluminum-based III–V nitride alloys, focusing on Al\textsubscript{1-x}Sc\textsubscript{x}N and Al\textsubscript{1-x}B\textsubscript{x}N. By adopting a unified 48-atom supercell framework, we were able to compare multiple competing polymorphs—wurtzite, rocksalt, zincblende, and hexagonal phases—on an equal footing, while systematically assessing the role of chemical disorder and exchange–correlation approximations.

Our results indicate that the stability range of the ferroelectric wurtzite phase is strongly underestimated by the virtual crystal approximation in Al\textsubscript{1-x}Sc\textsubscript{x}N and that, similarly, the VCA does not allow for strong distortions of the WZ phase in Al\textsubscript{1-x}B\textsubscript{x}N caused by the natural tendency of the B ion to be threefold coordinated; considering short-range chemical disorder such as in the SQS method is therefore critical. Evidently, deviation from perfect random alloying such as clustering, anticlustering or partial chemical ordering may influence the structural, ferroelectric and electronic properties of III-V ferroelectrics, as has been demonstrated in non-ferroelectric nitride~\cite{Bellaiche1998} or perovskite oxide~\cite{George2003, Prosandeev2016}  systems. This will be the focus of a follow-up study.

We also reveal a complex energetic landscape in Al\textsubscript{1-x}Sc\textsubscript{x}N, characterized by a competition between wurtzite, rocksalt, and hexagonal phases, including a near-degeneracy in the vicinity of the structural transition. This competition is reflected in the progressive reduction of spontaneous polarization with increasing scandium content, and highlights the delicate balance between coordination environment and polar distortion. The tendency of Al\textsubscript{1-x}Sc\textsubscript{x}N toward higher coordination structures (HX and RS) is likely driven by the increase in bonding channels provided by the hybridization of N $2p$ states with Sc $d$ states.  In contrast, Al\textsubscript{1-x}B\textsubscript{x}N exhibits markedly different behavior, driven by the smaller atomic size and bonding preferences of boron. We identify significant bond reconfiguration effects at intermediate concentrations, as well as a tendency toward lower coordination phases at high boron content. Notably, our calculations predict an enhancement of spontaneous polarization at low boron concentrations. 

Overall, our study provides a coherent and quantitatively grounded picture of alloying effects in III–V nitride ferroelectrics, while clarifying the quantitative effects of commonly used approximations for accurate first-principles predictions. These findings offer guidance for the design and optimization of nitride-based ferroelectric materials, and open perspectives for exploring more complex alloying strategies, strain engineering, and defect effects in future work.

\begin{acknowledgments}
This work is supported by a grant from the U.S. Department of Energy through grant agreement no. DE-SC0025479. Ab-initio calculations were performed on the Arkansas High Performance Computing Center, which is funded through multiple National Science Foundation grants and the Arkansas Economic Development Commission. This work was authored in part by the National Laboratory of the Rockies, operated by Alliance for Advanced Energy, LLC, for the U.S. Department of Energy (DOE) under contract no. DE-AC36-08GO28308. AZ and KY acknowledge funding from DOE Office of Science (SC), Basic Energy Sciences (BES), Materials Chemistry program. The views expressed in the article do not necessarily represent the views of the U.S. Department of Energy or the U.S. Government.
\end{acknowledgments}

\bibliography{biblio}

\begin{thebibliography}{70}%
\makeatletter
\providecommand \@ifxundefined [1]{%
 \@ifx{#1\undefined}
}%
\providecommand \@ifnum [1]{%
 \ifnum #1\expandafter \@firstoftwo
 \else \expandafter \@secondoftwo
 \fi
}%
\providecommand \@ifx [1]{%
 \ifx #1\expandafter \@firstoftwo
 \else \expandafter \@secondoftwo
 \fi
}%
\providecommand \natexlab [1]{#1}%
\providecommand \enquote  [1]{``#1''}%
\providecommand \bibnamefont  [1]{#1}%
\providecommand \bibfnamefont [1]{#1}%
\providecommand \citenamefont [1]{#1}%
\providecommand \href@noop [0]{\@secondoftwo}%
\providecommand \href [0]{\begingroup \@sanitize@url \@href}%
\providecommand \@href[1]{\@@startlink{#1}\@@href}%
\providecommand \@@href[1]{\endgroup#1\@@endlink}%
\providecommand \@sanitize@url [0]{\catcode `\\12\catcode `\$12\catcode
  `\&12\catcode `\#12\catcode `\^12\catcode `\_12\catcode `\%12\relax}%
\providecommand \@@startlink[1]{}%
\providecommand \@@endlink[0]{}%
\providecommand \url  [0]{\begingroup\@sanitize@url \@url }%
\providecommand \@url [1]{\endgroup\@href {#1}{\urlprefix }}%
\providecommand \urlprefix  [0]{URL }%
\providecommand \Eprint [0]{\href }%
\providecommand \doibase [0]{https://doi.org/}%
\providecommand \selectlanguage [0]{\@gobble}%
\providecommand \bibinfo  [0]{\@secondoftwo}%
\providecommand \bibfield  [0]{\@secondoftwo}%
\providecommand \translation [1]{[#1]}%
\providecommand \BibitemOpen [0]{}%
\providecommand \bibitemStop [0]{}%
\providecommand \bibitemNoStop [0]{.\EOS\space}%
\providecommand \EOS [0]{\spacefactor3000\relax}%
\providecommand \BibitemShut  [1]{\csname bibitem#1\endcsname}%
\let\auto@bib@innerbib\@empty
\bibitem [{\citenamefont {Fichtner}\ \emph {et~al.}(2019)\citenamefont
  {Fichtner}, \citenamefont {Wolff}, \citenamefont {Lofink}, \citenamefont
  {Kienle},\ and\ \citenamefont {Wagner}}]{Fichtner2019}%
  \BibitemOpen
  \bibfield  {author} {\bibinfo {author} {\bibfnamefont {S.}~\bibnamefont
  {Fichtner}}, \bibinfo {author} {\bibfnamefont {N.}~\bibnamefont {Wolff}},
  \bibinfo {author} {\bibfnamefont {F.}~\bibnamefont {Lofink}}, \bibinfo
  {author} {\bibfnamefont {L.}~\bibnamefont {Kienle}},\ and\ \bibinfo {author}
  {\bibfnamefont {B.}~\bibnamefont {Wagner}},\ }\bibfield  {title} {\bibinfo
  {title} {Alscn: A iii-v semiconductor based ferroelectric},\ }\href
  {https://doi.org/10.1063/1.5084945} {\bibfield  {journal} {\bibinfo
  {journal} {Journal of Applied Physics}\ }\textbf {\bibinfo {volume} {125}},\
  \bibinfo {pages} {114103} (\bibinfo {year} {2019})}\BibitemShut {NoStop}%
\bibitem [{\citenamefont {Islam}\ \emph {et~al.}(2021)\citenamefont {Islam},
  \citenamefont {Wolff}, \citenamefont {Yassine}, \citenamefont {Schönweger},
  \citenamefont {Christian}, \citenamefont {Kohlstedt}, \citenamefont
  {Ambacher}, \citenamefont {Lofink}, \citenamefont {Kienle},\ and\
  \citenamefont {Fichtner}}]{Islam2021}%
  \BibitemOpen
  \bibfield  {author} {\bibinfo {author} {\bibfnamefont {M.~R.}\ \bibnamefont
  {Islam}}, \bibinfo {author} {\bibfnamefont {N.}~\bibnamefont {Wolff}},
  \bibinfo {author} {\bibfnamefont {M.}~\bibnamefont {Yassine}}, \bibinfo
  {author} {\bibfnamefont {G.}~\bibnamefont {Schönweger}}, \bibinfo {author}
  {\bibfnamefont {B.}~\bibnamefont {Christian}}, \bibinfo {author}
  {\bibfnamefont {H.}~\bibnamefont {Kohlstedt}}, \bibinfo {author}
  {\bibfnamefont {O.}~\bibnamefont {Ambacher}}, \bibinfo {author}
  {\bibfnamefont {F.}~\bibnamefont {Lofink}}, \bibinfo {author} {\bibfnamefont
  {L.}~\bibnamefont {Kienle}},\ and\ \bibinfo {author} {\bibfnamefont
  {S.}~\bibnamefont {Fichtner}},\ }\bibfield  {title} {\bibinfo {title} {On the
  exceptional temperature stability of ferroelectric al1-xscxn thin films},\
  }\href {https://doi.org/10.1063/5.0053649} {\bibfield  {journal} {\bibinfo
  {journal} {Applied Physics Letters}\ }\textbf {\bibinfo {volume} {118}},\
  \bibinfo {pages} {232905} (\bibinfo {year} {2021})}\BibitemShut {NoStop}%
\bibitem [{\citenamefont {Kim}\ \emph {et~al.}(2024)\citenamefont {Kim},
  \citenamefont {Han}, \citenamefont {Zhang}, \citenamefont {Musavigharavi},
  \citenamefont {Zheng}, \citenamefont {Pradhan}, \citenamefont {Stach},
  \citenamefont {Olsson},\ and\ \citenamefont {Jariwala}}]{Kim2024}%
  \BibitemOpen
  \bibfield  {author} {\bibinfo {author} {\bibfnamefont {K.-H.}\ \bibnamefont
  {Kim}}, \bibinfo {author} {\bibfnamefont {Z.}~\bibnamefont {Han}}, \bibinfo
  {author} {\bibfnamefont {Y.}~\bibnamefont {Zhang}}, \bibinfo {author}
  {\bibfnamefont {P.}~\bibnamefont {Musavigharavi}}, \bibinfo {author}
  {\bibfnamefont {J.}~\bibnamefont {Zheng}}, \bibinfo {author} {\bibfnamefont
  {D.~K.}\ \bibnamefont {Pradhan}}, \bibinfo {author} {\bibfnamefont {E.~A.}\
  \bibnamefont {Stach}}, \bibinfo {author} {\bibfnamefont {R.~H.}\ \bibnamefont
  {Olsson}},\ and\ \bibinfo {author} {\bibfnamefont {D.}~\bibnamefont
  {Jariwala}},\ }\bibfield  {title} {\bibinfo {title} {Multistate, ultrathin,
  back-end-of-line-compatible alscn ferroelectric diodes},\ }\href
  {https://doi.org/10.1021/acsnano.4c03541} {\bibfield  {journal} {\bibinfo
  {journal} {ACS Nano}\ }\textbf {\bibinfo {volume} {18}},\ \bibinfo {pages}
  {15925} (\bibinfo {year} {2024})}\BibitemShut {NoStop}%
\bibitem [{\citenamefont {Hu}\ \emph {et~al.}(2025)\citenamefont {Hu},
  \citenamefont {Cho}, \citenamefont {Rai}, \citenamefont {Bao}, \citenamefont
  {Zhang}, \citenamefont {Qu}, \citenamefont {He}, \citenamefont {Ji},
  \citenamefont {Leblanc}, \citenamefont {Kim}, \citenamefont {Han},
  \citenamefont {Qiu}, \citenamefont {Du}, \citenamefont {Stach}, \citenamefont
  {Olsson},\ and\ \citenamefont {Jariwala}}]{Hu2025}%
  \BibitemOpen
  \bibfield  {author} {\bibinfo {author} {\bibfnamefont {Z.}~\bibnamefont
  {Hu}}, \bibinfo {author} {\bibfnamefont {H.}~\bibnamefont {Cho}}, \bibinfo
  {author} {\bibfnamefont {R.~K.}\ \bibnamefont {Rai}}, \bibinfo {author}
  {\bibfnamefont {K.}~\bibnamefont {Bao}}, \bibinfo {author} {\bibfnamefont
  {Y.}~\bibnamefont {Zhang}}, \bibinfo {author} {\bibfnamefont
  {Z.}~\bibnamefont {Qu}}, \bibinfo {author} {\bibfnamefont {Y.}~\bibnamefont
  {He}}, \bibinfo {author} {\bibfnamefont {Y.}~\bibnamefont {Ji}}, \bibinfo
  {author} {\bibfnamefont {C.}~\bibnamefont {Leblanc}}, \bibinfo {author}
  {\bibfnamefont {K.-H.}\ \bibnamefont {Kim}}, \bibinfo {author} {\bibfnamefont
  {Z.}~\bibnamefont {Han}}, \bibinfo {author} {\bibfnamefont {Z.}~\bibnamefont
  {Qiu}}, \bibinfo {author} {\bibfnamefont {X.}~\bibnamefont {Du}}, \bibinfo
  {author} {\bibfnamefont {E.~A.}\ \bibnamefont {Stach}}, \bibinfo {author}
  {\bibfnamefont {R.}~\bibnamefont {Olsson}},\ and\ \bibinfo {author}
  {\bibfnamefont {D.}~\bibnamefont {Jariwala}},\ }\bibfield  {title} {\bibinfo
  {title} {Demonstration of highly scaled alscn ferroelectric diode memory with
  a storage density of $>$ 100 mbit/mm\textsuperscript{2}},\ }\href
  {https://doi.org/10.1021/acs.nanolett.5c02961} {\bibfield  {journal}
  {\bibinfo  {journal} {Nano Letters}\ }\textbf {\bibinfo {volume} {25}},\
  \bibinfo {pages} {13748} (\bibinfo {year} {2025})}\BibitemShut {NoStop}%
\bibitem [{\citenamefont {Yang}\ \emph {et~al.}(2024)\citenamefont {Yang},
  \citenamefont {Wang}, \citenamefont {Mu}, \citenamefont {Xie}, \citenamefont
  {Wang}, \citenamefont {He}, \citenamefont {Shen}, \citenamefont {Liu},
  \citenamefont {de~Walle},\ and\ \citenamefont {Tang}}]{Yang2024}%
  \BibitemOpen
  \bibfield  {author} {\bibinfo {author} {\bibfnamefont {G.}~\bibnamefont
  {Yang}}, \bibinfo {author} {\bibfnamefont {H.}~\bibnamefont {Wang}}, \bibinfo
  {author} {\bibfnamefont {S.}~\bibnamefont {Mu}}, \bibinfo {author}
  {\bibfnamefont {H.}~\bibnamefont {Xie}}, \bibinfo {author} {\bibfnamefont
  {T.}~\bibnamefont {Wang}}, \bibinfo {author} {\bibfnamefont {C.}~\bibnamefont
  {He}}, \bibinfo {author} {\bibfnamefont {M.}~\bibnamefont {Shen}}, \bibinfo
  {author} {\bibfnamefont {M.}~\bibnamefont {Liu}}, \bibinfo {author}
  {\bibfnamefont {C.~G.~V.}\ \bibnamefont {de~Walle}},\ and\ \bibinfo {author}
  {\bibfnamefont {H.~X.}\ \bibnamefont {Tang}},\ }\bibfield  {title} {\bibinfo
  {title} {Unveiling the pockels coefficient of ferroelectric nitride scaln},\
  }\href {https://doi.org/10.1038/s41467-024-53895-x} {\bibfield  {journal}
  {\bibinfo  {journal} {Nature Communications}\ }\textbf {\bibinfo {volume}
  {15}},\ \bibinfo {pages} {9538} (\bibinfo {year} {2024})}\BibitemShut
  {NoStop}%
\bibitem [{\citenamefont {Wang}\ \emph {et~al.}(2025)\citenamefont {Wang},
  \citenamefont {Mu},\ and\ \citenamefont {de~Walle}}]{Wang2025_eo}%
  \BibitemOpen
  \bibfield  {author} {\bibinfo {author} {\bibfnamefont {H.}~\bibnamefont
  {Wang}}, \bibinfo {author} {\bibfnamefont {S.}~\bibnamefont {Mu}},\ and\
  \bibinfo {author} {\bibfnamefont {C.~G.~V.}\ \bibnamefont {de~Walle}},\
  }\bibfield  {title} {\bibinfo {title} {Towards higher electro-optic response
  in alscn},\ }\href {https://doi.org/10.1063/5.0244434} {\bibfield  {journal}
  {\bibinfo  {journal} {Applied Physics Letters}\ }\textbf {\bibinfo {volume}
  {126}},\ \bibinfo {pages} {041901} (\bibinfo {year} {2025})}\BibitemShut
  {NoStop}%
\bibitem [{\citenamefont {Wang}\ \emph {et~al.}(2024)\citenamefont {Wang},
  \citenamefont {Yang}, \citenamefont {Liu}, \citenamefont {Wang},\ and\
  \citenamefont {Mi}}]{Wang2024}%
  \BibitemOpen
  \bibfield  {author} {\bibinfo {author} {\bibfnamefont {D.}~\bibnamefont
  {Wang}}, \bibinfo {author} {\bibfnamefont {S.}~\bibnamefont {Yang}}, \bibinfo
  {author} {\bibfnamefont {J.}~\bibnamefont {Liu}}, \bibinfo {author}
  {\bibfnamefont {D.}~\bibnamefont {Wang}},\ and\ \bibinfo {author}
  {\bibfnamefont {Z.}~\bibnamefont {Mi}},\ }\bibfield  {title} {\bibinfo
  {title} {Perspectives on nitride ferroelectric semiconductors: Challenges and
  opportunities},\ }\href {https://doi.org/10.1063/5.0206005} {\bibfield
  {journal} {\bibinfo  {journal} {Applied Physics Letters}\ }\textbf {\bibinfo
  {volume} {124}},\ \bibinfo {pages} {150501} (\bibinfo {year}
  {2024})}\BibitemShut {NoStop}%
\bibitem [{\citenamefont {Farrer}\ and\ \citenamefont
  {Bellaiche}(2002)}]{Farrer2002}%
  \BibitemOpen
  \bibfield  {author} {\bibinfo {author} {\bibfnamefont {N.}~\bibnamefont
  {Farrer}}\ and\ \bibinfo {author} {\bibfnamefont {L.}~\bibnamefont
  {Bellaiche}},\ }\bibfield  {title} {\bibinfo {title} {Properties of hexagonal
  scn versus wurtzite gan and inn},\ }\href
  {https://doi.org/10.1103/PhysRevB.66.201203} {\bibfield  {journal} {\bibinfo
  {journal} {Physical Review B}\ }\textbf {\bibinfo {volume} {66}},\ \bibinfo
  {pages} {201203} (\bibinfo {year} {2002})}\BibitemShut {NoStop}%
\bibitem [{\citenamefont {Limpijumnong}\ and\ \citenamefont
  {Lambrecht}(2001)}]{Limpijumnong2001}%
  \BibitemOpen
  \bibfield  {author} {\bibinfo {author} {\bibfnamefont {S.}~\bibnamefont
  {Limpijumnong}}\ and\ \bibinfo {author} {\bibfnamefont {W.~R.~L.}\
  \bibnamefont {Lambrecht}},\ }\bibfield  {title} {\bibinfo {title}
  {Theoretical study of the relative stability of wurtzite and rocksalt phases
  in mgo and gan},\ }\href {https://doi.org/10.1103/PhysRevB.63.104103}
  {\bibfield  {journal} {\bibinfo  {journal} {Physical Review B}\ }\textbf
  {\bibinfo {volume} {63}},\ \bibinfo {pages} {104103} (\bibinfo {year}
  {2001})}\BibitemShut {NoStop}%
\bibitem [{\citenamefont {Pike}\ \emph {et~al.}(2025)\citenamefont {Pike},
  \citenamefont {Pachter}, \citenamefont {Kennedy},\ and\ \citenamefont
  {Glavin}}]{Pike2025}%
  \BibitemOpen
  \bibfield  {author} {\bibinfo {author} {\bibfnamefont {N.~A.}\ \bibnamefont
  {Pike}}, \bibinfo {author} {\bibfnamefont {R.}~\bibnamefont {Pachter}},
  \bibinfo {author} {\bibfnamefont {W.~J.}\ \bibnamefont {Kennedy}},\ and\
  \bibinfo {author} {\bibfnamefont {N.}~\bibnamefont {Glavin}},\ }\bibfield
  {title} {\bibinfo {title} {Understanding the structure and polarization in
  ferroelectric alscn, scgan, and alscgan from first principles calculations},\
  }\href {https://doi.org/10.1103/ngv8-vq2k} {\bibfield  {journal} {\bibinfo
  {journal} {Physical Review B}\ }\textbf {\bibinfo {volume} {112}},\ \bibinfo
  {pages} {085204} (\bibinfo {year} {2025})}\BibitemShut {NoStop}%
\bibitem [{\citenamefont {Calderon}\ \emph {et~al.}(2024)\citenamefont
  {Calderon}, \citenamefont {Hayden}, \citenamefont {Delower}, \citenamefont
  {Maria},\ and\ \citenamefont {Dickey}}]{CalderonV2024}%
  \BibitemOpen
  \bibfield  {author} {\bibinfo {author} {\bibfnamefont {S.}~\bibnamefont
  {Calderon}}, \bibinfo {author} {\bibfnamefont {J.}~\bibnamefont {Hayden}},
  \bibinfo {author} {\bibfnamefont {M.}~\bibnamefont {Delower}}, \bibinfo
  {author} {\bibfnamefont {J.~P.}\ \bibnamefont {Maria}},\ and\ \bibinfo
  {author} {\bibfnamefont {E.~C.}\ \bibnamefont {Dickey}},\ }\bibfield  {title}
  {\bibinfo {title} {Effect of boron concentration on local structure and
  spontaneous polarization in albn thin films},\ }\href
  {https://doi.org/10.1063/5.0179942} {\bibfield  {journal} {\bibinfo
  {journal} {APL Materials}\ }\textbf {\bibinfo {volume} {12}},\ \bibinfo
  {pages} {021105} (\bibinfo {year} {2024})}\BibitemShut {NoStop}%
\bibitem [{\citenamefont {Perdew}\ \emph {et~al.}(2008)\citenamefont {Perdew},
  \citenamefont {Ruzsinszky}, \citenamefont {Csonka}, \citenamefont {Vydrov},
  \citenamefont {Scuseria}, \citenamefont {Constantin}, \citenamefont {Zhou},\
  and\ \citenamefont {Burke}}]{Perdew2008}%
  \BibitemOpen
  \bibfield  {author} {\bibinfo {author} {\bibfnamefont {J.~P.}\ \bibnamefont
  {Perdew}}, \bibinfo {author} {\bibfnamefont {A.}~\bibnamefont {Ruzsinszky}},
  \bibinfo {author} {\bibfnamefont {G.~I.}\ \bibnamefont {Csonka}}, \bibinfo
  {author} {\bibfnamefont {O.~A.}\ \bibnamefont {Vydrov}}, \bibinfo {author}
  {\bibfnamefont {G.~E.}\ \bibnamefont {Scuseria}}, \bibinfo {author}
  {\bibfnamefont {L.~A.}\ \bibnamefont {Constantin}}, \bibinfo {author}
  {\bibfnamefont {X.}~\bibnamefont {Zhou}},\ and\ \bibinfo {author}
  {\bibfnamefont {K.}~\bibnamefont {Burke}},\ }\bibfield  {title} {\bibinfo
  {title} {Restoring the density-gradient expansion for exchange in solids and
  surfaces},\ }\href {https://doi.org/10.1103/PhysRevLett.100.136406}
  {\bibfield  {journal} {\bibinfo  {journal} {Physical Review Letters}\
  }\textbf {\bibinfo {volume} {100}},\ \bibinfo {pages} {136406} (\bibinfo
  {year} {2008})}\BibitemShut {NoStop}%
\bibitem [{\citenamefont {Perdew}\ \emph {et~al.}(1996)\citenamefont {Perdew},
  \citenamefont {Burke},\ and\ \citenamefont {Ernzerhof}}]{Perdew1996}%
  \BibitemOpen
  \bibfield  {author} {\bibinfo {author} {\bibfnamefont {J.~P.}\ \bibnamefont
  {Perdew}}, \bibinfo {author} {\bibfnamefont {K.}~\bibnamefont {Burke}},\ and\
  \bibinfo {author} {\bibfnamefont {M.}~\bibnamefont {Ernzerhof}},\ }\bibfield
  {title} {\bibinfo {title} {Generalized gradient approximation made simple},\
  }\href {https://doi.org/10.1103/PhysRevLett.77.3865} {\bibfield  {journal}
  {\bibinfo  {journal} {Physical Review Letters}\ }\textbf {\bibinfo {volume}
  {77}},\ \bibinfo {pages} {3865} (\bibinfo {year} {1996})}\BibitemShut
  {NoStop}%
\bibitem [{\citenamefont {Yazawa}\ \emph
  {et~al.}(2022{\natexlab{a}})\citenamefont {Yazawa}, \citenamefont {Mangum},
  \citenamefont {Gorai}, \citenamefont {Brennecka},\ and\ \citenamefont
  {Zakutayev}}]{Yazawa2022}%
  \BibitemOpen
  \bibfield  {author} {\bibinfo {author} {\bibfnamefont {K.}~\bibnamefont
  {Yazawa}}, \bibinfo {author} {\bibfnamefont {J.~S.}\ \bibnamefont {Mangum}},
  \bibinfo {author} {\bibfnamefont {P.}~\bibnamefont {Gorai}}, \bibinfo
  {author} {\bibfnamefont {G.~L.}\ \bibnamefont {Brennecka}},\ and\ \bibinfo
  {author} {\bibfnamefont {A.}~\bibnamefont {Zakutayev}},\ }\bibfield  {title}
  {\bibinfo {title} {Local chemical origin of ferroelectric behavior in
  wurtzite nitrides},\ }\href {https://doi.org/10.1039/D2TC02682A} {\bibfield
  {journal} {\bibinfo  {journal} {Journal of Materials Chemistry C}\ }\textbf
  {\bibinfo {volume} {10}},\ \bibinfo {pages} {17557} (\bibinfo {year}
  {2022}{\natexlab{a}})}\BibitemShut {NoStop}%
\bibitem [{\citenamefont {Calderon}\ \emph {et~al.}(2023)\citenamefont
  {Calderon}, \citenamefont {Hayden}, \citenamefont {Baksa}, \citenamefont
  {Tzou}, \citenamefont {Trolier-McKinstry}, \citenamefont {Dabo},
  \citenamefont {Maria},\ and\ \citenamefont {Dickey}}]{Calderon2023}%
  \BibitemOpen
  \bibfield  {author} {\bibinfo {author} {\bibfnamefont {S.}~\bibnamefont
  {Calderon}}, \bibinfo {author} {\bibfnamefont {J.}~\bibnamefont {Hayden}},
  \bibinfo {author} {\bibfnamefont {S.~M.}\ \bibnamefont {Baksa}}, \bibinfo
  {author} {\bibfnamefont {W.}~\bibnamefont {Tzou}}, \bibinfo {author}
  {\bibfnamefont {S.}~\bibnamefont {Trolier-McKinstry}}, \bibinfo {author}
  {\bibfnamefont {I.}~\bibnamefont {Dabo}}, \bibinfo {author} {\bibfnamefont
  {J.-P.}\ \bibnamefont {Maria}},\ and\ \bibinfo {author} {\bibfnamefont
  {E.~C.}\ \bibnamefont {Dickey}},\ }\bibfield  {title} {\bibinfo {title}
  {Atomic-scale polarization switching in wurtzite ferroelectrics},\ }\href
  {https://doi.org/10.1126/science.adh7670} {\bibfield  {journal} {\bibinfo
  {journal} {Science}\ }\textbf {\bibinfo {volume} {380}},\ \bibinfo {pages}
  {1034} (\bibinfo {year} {2023})}\BibitemShut {NoStop}%
\bibitem [{\citenamefont {Chen}\ \emph {et~al.}(2025)\citenamefont {Chen},
  \citenamefont {Wang}, \citenamefont {Tejerina}, \citenamefont {Yazawa},
  \citenamefont {Zakutayev}, \citenamefont {Paillard},\ and\ \citenamefont
  {Bellaiche}}]{Chen2025}%
  \BibitemOpen
  \bibfield  {author} {\bibinfo {author} {\bibfnamefont {P.}~\bibnamefont
  {Chen}}, \bibinfo {author} {\bibfnamefont {D.}~\bibnamefont {Wang}}, \bibinfo
  {author} {\bibfnamefont {A.~M.}\ \bibnamefont {Tejerina}}, \bibinfo {author}
  {\bibfnamefont {K.}~\bibnamefont {Yazawa}}, \bibinfo {author} {\bibfnamefont
  {A.}~\bibnamefont {Zakutayev}}, \bibinfo {author} {\bibfnamefont
  {C.}~\bibnamefont {Paillard}},\ and\ \bibinfo {author} {\bibfnamefont
  {L.}~\bibnamefont {Bellaiche}},\ }\bibfield  {title} {\bibinfo {title}
  {Towards a deeper fundamental understanding of (al,sc)n ferroelectric
  nitrides},\ }\href {https://doi.org/10.1103/9nv5-ryqr} {\bibfield  {journal}
  {\bibinfo  {journal} {Physical Review Materials}\ }\textbf {\bibinfo {volume}
  {9}},\ \bibinfo {pages} {124418} (\bibinfo {year} {2025})}\BibitemShut
  {NoStop}%
\bibitem [{\citenamefont {Arras}\ \emph {et~al.}(2026)\citenamefont {Arras},
  \citenamefont {Paillard},\ and\ \citenamefont {Bellaiche}}]{Arras2026}%
  \BibitemOpen
  \bibfield  {author} {\bibinfo {author} {\bibfnamefont {R.}~\bibnamefont
  {Arras}}, \bibinfo {author} {\bibfnamefont {C.}~\bibnamefont {Paillard}},\
  and\ \bibinfo {author} {\bibfnamefont {L.}~\bibnamefont {Bellaiche}},\
  }\bibfield  {title} {\bibinfo {title} {Magnetoelectric properties at the
  co/aln(0001) interface},\ }\href {https://doi.org/10.1103/vm5q-xbrm}
  {\bibfield  {journal} {\bibinfo  {journal} {Physical Review Materials}\
  }\textbf {\bibinfo {volume} {10}},\ \bibinfo {pages} {014411} (\bibinfo
  {year} {2026})}\BibitemShut {NoStop}%
\bibitem [{\citenamefont {Zhang}\ \emph {et~al.}(2017)\citenamefont {Zhang},
  \citenamefont {Sun}, \citenamefont {Perdew},\ and\ \citenamefont
  {Wu}}]{Zhang2017}%
  \BibitemOpen
  \bibfield  {author} {\bibinfo {author} {\bibfnamefont {Y.}~\bibnamefont
  {Zhang}}, \bibinfo {author} {\bibfnamefont {J.}~\bibnamefont {Sun}}, \bibinfo
  {author} {\bibfnamefont {J.~P.}\ \bibnamefont {Perdew}},\ and\ \bibinfo
  {author} {\bibfnamefont {X.}~\bibnamefont {Wu}},\ }\bibfield  {title}
  {\bibinfo {title} {Comparative first-principles studies of prototypical
  ferroelectric materials by lda, gga, and scan meta-gga},\ }\href
  {https://doi.org/10.1103/PhysRevB.96.035143} {\bibfield  {journal} {\bibinfo
  {journal} {Physical Review B}\ }\textbf {\bibinfo {volume} {96}},\ \bibinfo
  {pages} {035143} (\bibinfo {year} {2017})}\BibitemShut {NoStop}%
\bibitem [{\citenamefont {Momma}\ and\ \citenamefont
  {Izumi}(2008)}]{Momma2008}%
  \BibitemOpen
  \bibfield  {author} {\bibinfo {author} {\bibfnamefont {K.}~\bibnamefont
  {Momma}}\ and\ \bibinfo {author} {\bibfnamefont {F.}~\bibnamefont {Izumi}},\
  }\bibfield  {title} {\bibinfo {title} {Vesta : a three-dimensional
  visualization system for electronic and structural analysis},\ }\href
  {https://doi.org/10.1107/S0021889808012016} {\bibfield  {journal} {\bibinfo
  {journal} {Journal of Applied Crystallography}\ }\textbf {\bibinfo {volume}
  {41}},\ \bibinfo {pages} {653} (\bibinfo {year} {2008})}\BibitemShut
  {NoStop}%
\bibitem [{\citenamefont {Kresse}\ and\ \citenamefont
  {Hafner}(1994)}]{Kresse1994}%
  \BibitemOpen
  \bibfield  {author} {\bibinfo {author} {\bibfnamefont {G.}~\bibnamefont
  {Kresse}}\ and\ \bibinfo {author} {\bibfnamefont {J.}~\bibnamefont
  {Hafner}},\ }\bibfield  {title} {\bibinfo {title} {Ab initio
  molecular-dynamics simulation of the liquid-metal–amorphous-semiconductor
  transition in germanium},\ }\href {https://doi.org/10.1103/PhysRevB.49.14251}
  {\bibfield  {journal} {\bibinfo  {journal} {Physical Review B}\ }\textbf
  {\bibinfo {volume} {49}},\ \bibinfo {pages} {14251} (\bibinfo {year}
  {1994})}\BibitemShut {NoStop}%
\bibitem [{\citenamefont {Kresse}\ and\ \citenamefont
  {Furthmüller}(1996)}]{Kresse1996}%
  \BibitemOpen
  \bibfield  {author} {\bibinfo {author} {\bibfnamefont {G.}~\bibnamefont
  {Kresse}}\ and\ \bibinfo {author} {\bibfnamefont {J.}~\bibnamefont
  {Furthmüller}},\ }\bibfield  {title} {\bibinfo {title} {Efficiency of
  ab-initio total energy calculations for metals and semiconductors using a
  plane-wave basis set},\ }\href {https://doi.org/10.1016/0927-0256(96)00008-0}
  {\bibfield  {journal} {\bibinfo  {journal} {Computational Materials Science}\
  }\textbf {\bibinfo {volume} {6}},\ \bibinfo {pages} {15} (\bibinfo {year}
  {1996})}\BibitemShut {NoStop}%
\bibitem [{\citenamefont {Blöchl}(1994)}]{Blochl1994}%
  \BibitemOpen
  \bibfield  {author} {\bibinfo {author} {\bibfnamefont {P.~E.}\ \bibnamefont
  {Blöchl}},\ }\bibfield  {title} {\bibinfo {title} {Projector augmented-wave
  method},\ }\href {https://doi.org/10.1103/PhysRevB.50.17953} {\bibfield
  {journal} {\bibinfo  {journal} {Physical Review B}\ }\textbf {\bibinfo
  {volume} {50}},\ \bibinfo {pages} {17953} (\bibinfo {year}
  {1994})}\BibitemShut {NoStop}%
\bibitem [{\citenamefont {Kresse}\ and\ \citenamefont
  {Joubert}(1999)}]{Kresse1999}%
  \BibitemOpen
  \bibfield  {author} {\bibinfo {author} {\bibfnamefont {G.}~\bibnamefont
  {Kresse}}\ and\ \bibinfo {author} {\bibfnamefont {D.}~\bibnamefont
  {Joubert}},\ }\bibfield  {title} {\bibinfo {title} {From ultrasoft
  pseudopotentials to the projector augmented-wave method},\ }\href
  {https://doi.org/10.1103/PhysRevB.59.1758} {\bibfield  {journal} {\bibinfo
  {journal} {Physical Review B}\ }\textbf {\bibinfo {volume} {59}},\ \bibinfo
  {pages} {1758} (\bibinfo {year} {1999})}\BibitemShut {NoStop}%
\bibitem [{\citenamefont {Perdew}\ \emph {et~al.}(1997)\citenamefont {Perdew},
  \citenamefont {Burke},\ and\ \citenamefont {Ernzerhof}}]{Perdew1997}%
  \BibitemOpen
  \bibfield  {author} {\bibinfo {author} {\bibfnamefont {J.~P.}\ \bibnamefont
  {Perdew}}, \bibinfo {author} {\bibfnamefont {K.}~\bibnamefont {Burke}},\ and\
  \bibinfo {author} {\bibfnamefont {M.}~\bibnamefont {Ernzerhof}},\ }\bibfield
  {title} {\bibinfo {title} {Generalized gradient approximation made simple},\
  }\href {https://doi.org/10.1103/PhysRevLett.78.1396} {\bibfield  {journal}
  {\bibinfo  {journal} {Physical Review Letters}\ }\textbf {\bibinfo {volume}
  {78}},\ \bibinfo {pages} {1396} (\bibinfo {year} {1997})}\BibitemShut
  {NoStop}%
\bibitem [{\citenamefont {Sun}\ \emph {et~al.}(2015)\citenamefont {Sun},
  \citenamefont {Ruzsinszky},\ and\ \citenamefont {Perdew}}]{Sun2015}%
  \BibitemOpen
  \bibfield  {author} {\bibinfo {author} {\bibfnamefont {J.}~\bibnamefont
  {Sun}}, \bibinfo {author} {\bibfnamefont {A.}~\bibnamefont {Ruzsinszky}},\
  and\ \bibinfo {author} {\bibfnamefont {J.~P.}\ \bibnamefont {Perdew}},\
  }\bibfield  {title} {\bibinfo {title} {Strongly constrained and appropriately
  normed semilocal density functional},\ }\href
  {https://doi.org/10.1103/PhysRevLett.115.036402} {\bibfield  {journal}
  {\bibinfo  {journal} {Physical Review Letters}\ }\textbf {\bibinfo {volume}
  {115}},\ \bibinfo {pages} {036402} (\bibinfo {year} {2015})}\BibitemShut
  {NoStop}%
\bibitem [{\citenamefont {Sabatini}\ \emph {et~al.}(2013)\citenamefont
  {Sabatini}, \citenamefont {Gorni},\ and\ \citenamefont
  {de~Gironcoli}}]{Sabatini2013}%
  \BibitemOpen
  \bibfield  {author} {\bibinfo {author} {\bibfnamefont {R.}~\bibnamefont
  {Sabatini}}, \bibinfo {author} {\bibfnamefont {T.}~\bibnamefont {Gorni}},\
  and\ \bibinfo {author} {\bibfnamefont {S.}~\bibnamefont {de~Gironcoli}},\
  }\bibfield  {title} {\bibinfo {title} {Nonlocal van der waals density
  functional made simple and efficient},\ }\href
  {https://doi.org/10.1103/PhysRevB.87.041108} {\bibfield  {journal} {\bibinfo
  {journal} {Physical Review B}\ }\textbf {\bibinfo {volume} {87}},\ \bibinfo
  {pages} {041108} (\bibinfo {year} {2013})}\BibitemShut {NoStop}%
\bibitem [{\citenamefont {Peng}\ \emph {et~al.}(2016)\citenamefont {Peng},
  \citenamefont {Yang}, \citenamefont {Perdew},\ and\ \citenamefont
  {Sun}}]{Peng2016}%
  \BibitemOpen
  \bibfield  {author} {\bibinfo {author} {\bibfnamefont {H.}~\bibnamefont
  {Peng}}, \bibinfo {author} {\bibfnamefont {Z.-H.}\ \bibnamefont {Yang}},
  \bibinfo {author} {\bibfnamefont {J.~P.}\ \bibnamefont {Perdew}},\ and\
  \bibinfo {author} {\bibfnamefont {J.}~\bibnamefont {Sun}},\ }\bibfield
  {title} {\bibinfo {title} {Versatile van der waals density functional based
  on a meta-generalized gradient approximation},\ }\href
  {https://doi.org/10.1103/PhysRevX.6.041005} {\bibfield  {journal} {\bibinfo
  {journal} {Physical Review X}\ }\textbf {\bibinfo {volume} {6}},\ \bibinfo
  {pages} {041005} (\bibinfo {year} {2016})}\BibitemShut {NoStop}%
\bibitem [{\citenamefont {Zunger}\ \emph {et~al.}(1990)\citenamefont {Zunger},
  \citenamefont {Wei}, \citenamefont {Ferreira},\ and\ \citenamefont
  {Bernard}}]{Zunger1990}%
  \BibitemOpen
  \bibfield  {author} {\bibinfo {author} {\bibfnamefont {A.}~\bibnamefont
  {Zunger}}, \bibinfo {author} {\bibfnamefont {S.~H.}\ \bibnamefont {Wei}},
  \bibinfo {author} {\bibfnamefont {L.~G.}\ \bibnamefont {Ferreira}},\ and\
  \bibinfo {author} {\bibfnamefont {J.~E.}\ \bibnamefont {Bernard}},\
  }\bibfield  {title} {\bibinfo {title} {Special quasirandom structures},\
  }\href {https://doi.org/10.1103/PhysRevLett.65.353} {\bibfield  {journal}
  {\bibinfo  {journal} {Physical Review Letters}\ }\textbf {\bibinfo {volume}
  {65}},\ \bibinfo {pages} {353} (\bibinfo {year} {1990})}\BibitemShut
  {NoStop}%
\bibitem [{\citenamefont {van~de Walle}\ \emph {et~al.}(2013)\citenamefont
  {van~de Walle}, \citenamefont {Tiwary}, \citenamefont {de~Jong},
  \citenamefont {Olmsted}, \citenamefont {Asta}, \citenamefont {Dick},
  \citenamefont {Shin}, \citenamefont {Wang}, \citenamefont {Chen},\ and\
  \citenamefont {Liu}}]{vandeWalle2013}%
  \BibitemOpen
  \bibfield  {author} {\bibinfo {author} {\bibfnamefont {A.}~\bibnamefont
  {van~de Walle}}, \bibinfo {author} {\bibfnamefont {P.}~\bibnamefont
  {Tiwary}}, \bibinfo {author} {\bibfnamefont {M.}~\bibnamefont {de~Jong}},
  \bibinfo {author} {\bibfnamefont {D.}~\bibnamefont {Olmsted}}, \bibinfo
  {author} {\bibfnamefont {M.}~\bibnamefont {Asta}}, \bibinfo {author}
  {\bibfnamefont {A.}~\bibnamefont {Dick}}, \bibinfo {author} {\bibfnamefont
  {D.}~\bibnamefont {Shin}}, \bibinfo {author} {\bibfnamefont {Y.}~\bibnamefont
  {Wang}}, \bibinfo {author} {\bibfnamefont {L.-Q.}\ \bibnamefont {Chen}},\
  and\ \bibinfo {author} {\bibfnamefont {Z.-K.}\ \bibnamefont {Liu}},\
  }\bibfield  {title} {\bibinfo {title} {Efficient stochastic generation of
  special quasirandom structures},\ }\href
  {https://doi.org/10.1016/j.calphad.2013.06.006} {\bibfield  {journal}
  {\bibinfo  {journal} {Calphad}\ }\textbf {\bibinfo {volume} {42}},\ \bibinfo
  {pages} {13} (\bibinfo {year} {2013})}\BibitemShut {NoStop}%
\bibitem [{\citenamefont {Ghosez}\ \emph {et~al.}(2000)\citenamefont {Ghosez},
  \citenamefont {Dequesnes}, \citenamefont {Gonze},\ and\ \citenamefont
  {Rabe}}]{Ghosez2000}%
  \BibitemOpen
  \bibfield  {author} {\bibinfo {author} {\bibfnamefont {P.}~\bibnamefont
  {Ghosez}}, \bibinfo {author} {\bibfnamefont {D.}~\bibnamefont {Dequesnes}},
  \bibinfo {author} {\bibfnamefont {X.}~\bibnamefont {Gonze}},\ and\ \bibinfo
  {author} {\bibfnamefont {K.}~\bibnamefont {Rabe}},\ }\bibfield  {title}
  {\bibinfo {title} {First-principles study of lattice instabilities in
  ba\textsubscript{x}sr\textsubscript{1-x}tio\textsubscript{3}},\ }in\ \href
  {https://doi.org/10.1063/1.1324445} {\emph {\bibinfo {booktitle} {AIP
  Conference Proceedings}}},\ Vol.\ \bibinfo {volume} {535}\ (\bibinfo
  {publisher} {AIP},\ \bibinfo {year} {2000})\ pp.\ \bibinfo {pages}
  {102--110}\BibitemShut {NoStop}%
\bibitem [{\citenamefont {Bellaiche}\ and\ \citenamefont
  {Vanderbilt}(2000)}]{Bellaiche2000}%
  \BibitemOpen
  \bibfield  {author} {\bibinfo {author} {\bibfnamefont {L.}~\bibnamefont
  {Bellaiche}}\ and\ \bibinfo {author} {\bibfnamefont {D.}~\bibnamefont
  {Vanderbilt}},\ }\bibfield  {title} {\bibinfo {title} {Virtual crystal
  approximation revisited: Application to dielectric and piezoelectric
  properties of perovskites},\ }\href
  {https://doi.org/10.1103/PhysRevB.61.7877} {\bibfield  {journal} {\bibinfo
  {journal} {Physical Review B}\ }\textbf {\bibinfo {volume} {61}},\ \bibinfo
  {pages} {7877} (\bibinfo {year} {2000})}\BibitemShut {NoStop}%
\bibitem [{SM()}]{SM}%
  \BibitemOpen
  \href@noop {} {}\bibinfo {howpublished} {See Supplemental Material at
  \url{URL_will_be_inserted_by_publisher} for additional data on the influence
  of norm-conserved versus PAW pseudopotentials, the electronic bandgap of
  (Al,Sc)N and (Al,B)N, bowing parameters in the wurtzite and zincblende phases
  of (Al,B)N, with complementary
  references~\cite{mourad2012theory}\&\cite{Roy2024}.}\BibitemShut {Stop}%
\bibitem [{\citenamefont {Pease}(1950)}]{Pease1950}%
  \BibitemOpen
  \bibfield  {author} {\bibinfo {author} {\bibfnamefont {R.~S.}\ \bibnamefont
  {Pease}},\ }\bibfield  {title} {\bibinfo {title} {Crystal structure of boron
  nitride},\ }\href {https://doi.org/10.1038/165722b0} {\bibfield  {journal}
  {\bibinfo  {journal} {Nature}\ }\textbf {\bibinfo {volume} {165}},\ \bibinfo
  {pages} {722} (\bibinfo {year} {1950})}\BibitemShut {NoStop}%
\bibitem [{\citenamefont {Corrigan}\ and\ \citenamefont
  {Bundy}(1975)}]{Corrigan1975}%
  \BibitemOpen
  \bibfield  {author} {\bibinfo {author} {\bibfnamefont {F.~R.}\ \bibnamefont
  {Corrigan}}\ and\ \bibinfo {author} {\bibfnamefont {F.~P.}\ \bibnamefont
  {Bundy}},\ }\bibfield  {title} {\bibinfo {title} {Direct transitions among
  the allotropic forms of boron nitride at high pressures and temperatures},\
  }\href {https://doi.org/10.1063/1.431874} {\bibfield  {journal} {\bibinfo
  {journal} {The Journal of Chemical Physics}\ }\textbf {\bibinfo {volume}
  {63}},\ \bibinfo {pages} {3812} (\bibinfo {year} {1975})}\BibitemShut
  {NoStop}%
\bibitem [{\citenamefont {Cazorla}\ and\ \citenamefont
  {Gould}(2019)}]{Cazorla2019}%
  \BibitemOpen
  \bibfield  {author} {\bibinfo {author} {\bibfnamefont {C.}~\bibnamefont
  {Cazorla}}\ and\ \bibinfo {author} {\bibfnamefont {T.}~\bibnamefont
  {Gould}},\ }\bibfield  {title} {\bibinfo {title} {Polymorphism of bulk boron
  nitride},\ }\href {https://doi.org/10.1126/sciadv.aau5832} {\bibfield
  {journal} {\bibinfo  {journal} {Science Advances}\ }\textbf {\bibinfo
  {volume} {5}},\ \bibinfo {pages} {eaau5832} (\bibinfo {year}
  {2019})}\BibitemShut {NoStop}%
\bibitem [{\citenamefont {Hayden}\ \emph {et~al.}(2021)\citenamefont {Hayden},
  \citenamefont {Hossain}, \citenamefont {Xiong}, \citenamefont {Ferri},
  \citenamefont {Zhu}, \citenamefont {Imperatore}, \citenamefont {Giebink},
  \citenamefont {Trolier-Mckinstry}, \citenamefont {Dabo},\ and\ \citenamefont
  {Maria}}]{Hayden2021}%
  \BibitemOpen
  \bibfield  {author} {\bibinfo {author} {\bibfnamefont {J.}~\bibnamefont
  {Hayden}}, \bibinfo {author} {\bibfnamefont {M.~D.}\ \bibnamefont {Hossain}},
  \bibinfo {author} {\bibfnamefont {Y.}~\bibnamefont {Xiong}}, \bibinfo
  {author} {\bibfnamefont {K.}~\bibnamefont {Ferri}}, \bibinfo {author}
  {\bibfnamefont {W.}~\bibnamefont {Zhu}}, \bibinfo {author} {\bibfnamefont
  {M.~V.}\ \bibnamefont {Imperatore}}, \bibinfo {author} {\bibfnamefont
  {N.}~\bibnamefont {Giebink}}, \bibinfo {author} {\bibfnamefont
  {S.}~\bibnamefont {Trolier-Mckinstry}}, \bibinfo {author} {\bibfnamefont
  {I.}~\bibnamefont {Dabo}},\ and\ \bibinfo {author} {\bibfnamefont {J.~P.}\
  \bibnamefont {Maria}},\ }\bibfield  {title} {\bibinfo {title}
  {Ferroelectricity in boron-substituted aluminum nitride thin films},\ }\href
  {https://doi.org/10.1103/PhysRevMaterials.5.044412} {\bibfield  {journal}
  {\bibinfo  {journal} {Physical Review Materials}\ }\textbf {\bibinfo {volume}
  {5}},\ \bibinfo {pages} {044412} (\bibinfo {year} {2021})}\BibitemShut
  {NoStop}%
\bibitem [{\citenamefont {Csonka}\ \emph {et~al.}(2009)\citenamefont {Csonka},
  \citenamefont {Perdew}, \citenamefont {Ruzsinszky}, \citenamefont
  {Philipsen}, \citenamefont {Leb{\`e}gue}, \citenamefont {Paier},
  \citenamefont {Vydrov},\ and\ \citenamefont {{\'A}ngy{\'a}n}}]{Czonka2009}%
  \BibitemOpen
  \bibfield  {author} {\bibinfo {author} {\bibfnamefont {G.~I.}\ \bibnamefont
  {Csonka}}, \bibinfo {author} {\bibfnamefont {J.~P.}\ \bibnamefont {Perdew}},
  \bibinfo {author} {\bibfnamefont {A.}~\bibnamefont {Ruzsinszky}}, \bibinfo
  {author} {\bibfnamefont {P.~H.~T.}\ \bibnamefont {Philipsen}}, \bibinfo
  {author} {\bibfnamefont {S.}~\bibnamefont {Leb{\`e}gue}}, \bibinfo {author}
  {\bibfnamefont {J.}~\bibnamefont {Paier}}, \bibinfo {author} {\bibfnamefont
  {O.~A.}\ \bibnamefont {Vydrov}},\ and\ \bibinfo {author} {\bibfnamefont
  {J.~G.}\ \bibnamefont {{\'A}ngy{\'a}n}},\ }\bibfield  {title} {\bibinfo
  {title} {Assessing the performance of recent density functionals for bulk
  solids},\ }\href {https://doi.org/10.1103/PhysRevB.79.155107} {\bibfield
  {journal} {\bibinfo  {journal} {Physical Review B}\ }\textbf {\bibinfo
  {volume} {79}},\ \bibinfo {pages} {155107} (\bibinfo {year}
  {2009})}\BibitemShut {NoStop}%
\bibitem [{\citenamefont {Togo}(2023)}]{phonopy-phono3py-JPSJ}%
  \BibitemOpen
  \bibfield  {author} {\bibinfo {author} {\bibfnamefont {A.}~\bibnamefont
  {Togo}},\ }\bibfield  {title} {\bibinfo {title} {First-principles phonon
  calculations with phonopy and phono3py},\ }\href
  {https://doi.org/10.7566/JPSJ.92.012001} {\bibfield  {journal} {\bibinfo
  {journal} {Journal of the Physical Society of Japan}\ }\textbf {\bibinfo
  {volume} {92}},\ \bibinfo {pages} {012001} (\bibinfo {year}
  {2023})}\BibitemShut {NoStop}%
\bibitem [{\citenamefont {Talley}\ \emph {et~al.}(2018)\citenamefont {Talley},
  \citenamefont {Millican}, \citenamefont {Mangum}, \citenamefont {Siol},
  \citenamefont {Musgrave}, \citenamefont {Gorman}, \citenamefont {Holder},
  \citenamefont {Zakutayev},\ and\ \citenamefont {Brennecka}}]{Talley2018}%
  \BibitemOpen
  \bibfield  {author} {\bibinfo {author} {\bibfnamefont {K.~R.}\ \bibnamefont
  {Talley}}, \bibinfo {author} {\bibfnamefont {S.~L.}\ \bibnamefont
  {Millican}}, \bibinfo {author} {\bibfnamefont {J.}~\bibnamefont {Mangum}},
  \bibinfo {author} {\bibfnamefont {S.}~\bibnamefont {Siol}}, \bibinfo {author}
  {\bibfnamefont {C.~B.}\ \bibnamefont {Musgrave}}, \bibinfo {author}
  {\bibfnamefont {B.}~\bibnamefont {Gorman}}, \bibinfo {author} {\bibfnamefont
  {A.~M.}\ \bibnamefont {Holder}}, \bibinfo {author} {\bibfnamefont
  {A.}~\bibnamefont {Zakutayev}},\ and\ \bibinfo {author} {\bibfnamefont
  {G.~L.}\ \bibnamefont {Brennecka}},\ }\bibfield  {title} {\bibinfo {title}
  {Implications of heterostructural alloying for enhanced piezoelectric
  performance of (al,sc)n},\ }\href
  {https://doi.org/10.1103/PhysRevMaterials.2.063802} {\bibfield  {journal}
  {\bibinfo  {journal} {Physical Review Materials}\ }\textbf {\bibinfo {volume}
  {2}},\ \bibinfo {pages} {063802} (\bibinfo {year} {2018})}\BibitemShut
  {NoStop}%
\bibitem [{\citenamefont {Paul}\ \emph {et~al.}(2017)\citenamefont {Paul},
  \citenamefont {Sun}, \citenamefont {Perdew},\ and\ \citenamefont
  {Waghmare}}]{Paul2017}%
  \BibitemOpen
  \bibfield  {author} {\bibinfo {author} {\bibfnamefont {A.}~\bibnamefont
  {Paul}}, \bibinfo {author} {\bibfnamefont {J.}~\bibnamefont {Sun}}, \bibinfo
  {author} {\bibfnamefont {J.~P.}\ \bibnamefont {Perdew}},\ and\ \bibinfo
  {author} {\bibfnamefont {U.~V.}\ \bibnamefont {Waghmare}},\ }\bibfield
  {title} {\bibinfo {title} {Accuracy of first-principles interatomic
  interactions and predictions of ferroelectric phase transitions in perovskite
  oxides: Energy functional and effective hamiltonian},\ }\href
  {https://doi.org/10.1103/PhysRevB.95.054111} {\bibfield  {journal} {\bibinfo
  {journal} {Physical Review B}\ }\textbf {\bibinfo {volume} {95}},\ \bibinfo
  {pages} {054111} (\bibinfo {year} {2017})}\BibitemShut {NoStop}%
\bibitem [{\citenamefont {Jiang}\ \emph {et~al.}(2024)\citenamefont {Jiang},
  \citenamefont {Zhang}, \citenamefont {Paillard}, \citenamefont {Xiang},\ and\
  \citenamefont {Bellaiche}}]{PhysRevB.110.054101}%
  \BibitemOpen
  \bibfield  {author} {\bibinfo {author} {\bibfnamefont {Z.}~\bibnamefont
  {Jiang}}, \bibinfo {author} {\bibfnamefont {Z.}~\bibnamefont {Zhang}},
  \bibinfo {author} {\bibfnamefont {C.}~\bibnamefont {Paillard}}, \bibinfo
  {author} {\bibfnamefont {H.}~\bibnamefont {Xiang}},\ and\ \bibinfo {author}
  {\bibfnamefont {L.}~\bibnamefont {Bellaiche}},\ }\bibfield  {title} {\bibinfo
  {title} {Strain-induced bent domains in ferroelectric nitrides},\ }\href
  {https://doi.org/10.1103/PhysRevB.110.054101} {\bibfield  {journal} {\bibinfo
   {journal} {Physical Review B}\ }\textbf {\bibinfo {volume} {110}},\ \bibinfo
  {pages} {054101} (\bibinfo {year} {2024})}\BibitemShut {NoStop}%
\bibitem [{\citenamefont {Yazawa}\ \emph
  {et~al.}(2022{\natexlab{b}})\citenamefont {Yazawa}, \citenamefont
  {Zakutayev},\ and\ \citenamefont {Brennecka}}]{Yazawa2022b}%
  \BibitemOpen
  \bibfield  {author} {\bibinfo {author} {\bibfnamefont {K.}~\bibnamefont
  {Yazawa}}, \bibinfo {author} {\bibfnamefont {A.}~\bibnamefont {Zakutayev}},\
  and\ \bibinfo {author} {\bibfnamefont {G.~L.}\ \bibnamefont {Brennecka}},\
  }\bibfield  {title} {\bibinfo {title} {A landau–devonshire analysis of
  strain effects on ferroelectric al\textsubscript{1-x}sc\textsubscript{x}n},\
  }\bibfield  {journal} {\bibinfo  {journal} {Applied Physics Letters}\
  }\textbf {\bibinfo {volume} {121}},\ \href
  {https://doi.org/10.1063/5.0098979} {10.1063/5.0098979} (\bibinfo {year}
  {2022}{\natexlab{b}})\BibitemShut {NoStop}%
\bibitem [{\citenamefont {Hirata}\ \emph {et~al.}(2025)\citenamefont {Hirata},
  \citenamefont {Niitsu}, \citenamefont {Anggraini}, \citenamefont {Kageura},
  \citenamefont {Uehara}, \citenamefont {Yamada},\ and\ \citenamefont
  {Akiyama}}]{Hirata2025}%
  \BibitemOpen
  \bibfield  {author} {\bibinfo {author} {\bibfnamefont {K.}~\bibnamefont
  {Hirata}}, \bibinfo {author} {\bibfnamefont {K.}~\bibnamefont {Niitsu}},
  \bibinfo {author} {\bibfnamefont {S.~A.}\ \bibnamefont {Anggraini}}, \bibinfo
  {author} {\bibfnamefont {T.}~\bibnamefont {Kageura}}, \bibinfo {author}
  {\bibfnamefont {M.}~\bibnamefont {Uehara}}, \bibinfo {author} {\bibfnamefont
  {H.}~\bibnamefont {Yamada}},\ and\ \bibinfo {author} {\bibfnamefont
  {M.}~\bibnamefont {Akiyama}},\ }\bibfield  {title} {\bibinfo {title}
  {Enhancing the piezoelectric performance of nitride thin films through
  interfacial engineering},\ }\href
  {https://doi.org/10.1016/j.mattod.2024.12.011} {\bibfield  {journal}
  {\bibinfo  {journal} {Materials Today}\ }\textbf {\bibinfo {volume} {83}},\
  \bibinfo {pages} {85} (\bibinfo {year} {2025})}\BibitemShut {NoStop}%
\bibitem [{\citenamefont {Liu}\ \emph {et~al.}(2026)\citenamefont {Liu},
  \citenamefont {Cohen}, \citenamefont {Neufeld},\ and\ \citenamefont
  {Chen}}]{Liu2026}%
  \BibitemOpen
  \bibfield  {author} {\bibinfo {author} {\bibfnamefont {S.}~\bibnamefont
  {Liu}}, \bibinfo {author} {\bibfnamefont {O.}~\bibnamefont {Cohen}}, \bibinfo
  {author} {\bibfnamefont {O.}~\bibnamefont {Neufeld}},\ and\ \bibinfo {author}
  {\bibfnamefont {P.}~\bibnamefont {Chen}},\ }\bibfield  {title} {\bibinfo
  {title} {Laser-driven structural transformation from a bulk crystal to a
  layered material},\ }\href {https://doi.org/10.1103/fv5l-rjyv} {\bibfield
  {journal} {\bibinfo  {journal} {Physical Review Letters}\ }\textbf {\bibinfo
  {volume} {136}},\ \bibinfo {pages} {086902} (\bibinfo {year}
  {2026})}\BibitemShut {NoStop}%
\bibitem [{\citenamefont {Solozhenko}\ \emph {et~al.}(1999)\citenamefont
  {Solozhenko}, \citenamefont {Turkevich},\ and\ \citenamefont
  {Holzapfel}}]{Solozhenko1999}%
  \BibitemOpen
  \bibfield  {author} {\bibinfo {author} {\bibfnamefont {V.~L.}\ \bibnamefont
  {Solozhenko}}, \bibinfo {author} {\bibfnamefont {V.~Z.}\ \bibnamefont
  {Turkevich}},\ and\ \bibinfo {author} {\bibfnamefont {W.~B.}\ \bibnamefont
  {Holzapfel}},\ }\bibfield  {title} {\bibinfo {title} {Refined phase diagram
  of boron nitride},\ }\href {https://doi.org/10.1021/jp984682c} {\bibfield
  {journal} {\bibinfo  {journal} {The Journal of Physical Chemistry B}\
  }\textbf {\bibinfo {volume} {103}},\ \bibinfo {pages} {2903} (\bibinfo {year}
  {1999})}\BibitemShut {NoStop}%
\bibitem [{\citenamefont {Satoh}\ \emph {et~al.}(2022)\citenamefont {Satoh},
  \citenamefont {Ohtaka}, \citenamefont {Shimatsu},\ and\ \citenamefont
  {Tanaka}}]{Satoh2022}%
  \BibitemOpen
  \bibfield  {author} {\bibinfo {author} {\bibfnamefont {S.}~\bibnamefont
  {Satoh}}, \bibinfo {author} {\bibfnamefont {K.}~\bibnamefont {Ohtaka}},
  \bibinfo {author} {\bibfnamefont {T.}~\bibnamefont {Shimatsu}},\ and\
  \bibinfo {author} {\bibfnamefont {S.}~\bibnamefont {Tanaka}},\ }\bibfield
  {title} {\bibinfo {title} {Crystal structure deformation and phase transition
  of alscn thin films in whole sc concentration range},\ }\href
  {https://doi.org/10.1063/5.0087505} {\bibfield  {journal} {\bibinfo
  {journal} {Journal of Applied Physics}\ }\textbf {\bibinfo {volume} {132}},\
  \bibinfo {pages} {025103} (\bibinfo {year} {2022})}\BibitemShut {NoStop}%
\bibitem [{\citenamefont {Schulz}\ and\ \citenamefont
  {Thiemann}(1977)}]{Schulz1977}%
  \BibitemOpen
  \bibfield  {author} {\bibinfo {author} {\bibfnamefont {H.}~\bibnamefont
  {Schulz}}\ and\ \bibinfo {author} {\bibfnamefont {K.}~\bibnamefont
  {Thiemann}},\ }\bibfield  {title} {\bibinfo {title} {Crystal structure
  refinement of aln and gan},\ }\href
  {https://doi.org/10.1016/0038-1098(77)90959-0} {\bibfield  {journal}
  {\bibinfo  {journal} {Solid State Communications}\ }\textbf {\bibinfo
  {volume} {23}},\ \bibinfo {pages} {815} (\bibinfo {year} {1977})}\BibitemShut
  {NoStop}%
\bibitem [{\citenamefont {Akiyama}\ \emph {et~al.}(2009)\citenamefont
  {Akiyama}, \citenamefont {Kamohara}, \citenamefont {Kano}, \citenamefont
  {Teshigahara}, \citenamefont {Takeuchi},\ and\ \citenamefont
  {Kawahara}}]{Akiyama2009}%
  \BibitemOpen
  \bibfield  {author} {\bibinfo {author} {\bibfnamefont {M.}~\bibnamefont
  {Akiyama}}, \bibinfo {author} {\bibfnamefont {T.}~\bibnamefont {Kamohara}},
  \bibinfo {author} {\bibfnamefont {K.}~\bibnamefont {Kano}}, \bibinfo {author}
  {\bibfnamefont {A.}~\bibnamefont {Teshigahara}}, \bibinfo {author}
  {\bibfnamefont {Y.}~\bibnamefont {Takeuchi}},\ and\ \bibinfo {author}
  {\bibfnamefont {N.}~\bibnamefont {Kawahara}},\ }\bibfield  {title} {\bibinfo
  {title} {Enhancement of piezoelectric response in scandium aluminum nitride
  alloy thin films prepared by dual reactive cosputtering},\ }\href
  {https://doi.org/10.1002/adma.200802611} {\bibfield  {journal} {\bibinfo
  {journal} {Advanced Materials}\ }\textbf {\bibinfo {volume} {21}},\ \bibinfo
  {pages} {593} (\bibinfo {year} {2009})}\BibitemShut {NoStop}%
\bibitem [{\citenamefont {Alvarez}\ \emph {et~al.}(2023)\citenamefont
  {Alvarez}, \citenamefont {Casamento}, \citenamefont {van Deurzen},
  \citenamefont {Khan}, \citenamefont {Khan}, \citenamefont {Jeong},
  \citenamefont {Ahmadi}, \citenamefont {Xing}, \citenamefont {Jena},\ and\
  \citenamefont {Tian}}]{Alvarez2023}%
  \BibitemOpen
  \bibfield  {author} {\bibinfo {author} {\bibfnamefont {G.~A.}\ \bibnamefont
  {Alvarez}}, \bibinfo {author} {\bibfnamefont {J.}~\bibnamefont {Casamento}},
  \bibinfo {author} {\bibfnamefont {L.}~\bibnamefont {van Deurzen}}, \bibinfo
  {author} {\bibfnamefont {M.~I.}\ \bibnamefont {Khan}}, \bibinfo {author}
  {\bibfnamefont {K.}~\bibnamefont {Khan}}, \bibinfo {author} {\bibfnamefont
  {E.}~\bibnamefont {Jeong}}, \bibinfo {author} {\bibfnamefont
  {E.}~\bibnamefont {Ahmadi}}, \bibinfo {author} {\bibfnamefont {H.~G.}\
  \bibnamefont {Xing}}, \bibinfo {author} {\bibfnamefont {D.}~\bibnamefont
  {Jena}},\ and\ \bibinfo {author} {\bibfnamefont {Z.}~\bibnamefont {Tian}},\
  }\bibfield  {title} {\bibinfo {title} {Thermal conductivity enhancement of
  aluminum scandium nitride grown by molecular beam epitaxy},\ }\href
  {https://doi.org/10.1080/21663831.2023.2279667} {\bibfield  {journal}
  {\bibinfo  {journal} {Materials Research Letters}\ }\textbf {\bibinfo
  {volume} {11}},\ \bibinfo {pages} {1048} (\bibinfo {year}
  {2023})}\BibitemShut {NoStop}%
\bibitem [{\citenamefont {Cordero}\ \emph {et~al.}(2008)\citenamefont
  {Cordero}, \citenamefont {Gómez}, \citenamefont {Platero-Prats},
  \citenamefont {Revés}, \citenamefont {Echeverría}, \citenamefont
  {Cremades}, \citenamefont {Barragán},\ and\ \citenamefont
  {Alvarez}}]{Cordero2008}%
  \BibitemOpen
  \bibfield  {author} {\bibinfo {author} {\bibfnamefont {B.}~\bibnamefont
  {Cordero}}, \bibinfo {author} {\bibfnamefont {V.}~\bibnamefont {Gómez}},
  \bibinfo {author} {\bibfnamefont {A.~E.}\ \bibnamefont {Platero-Prats}},
  \bibinfo {author} {\bibfnamefont {M.}~\bibnamefont {Revés}}, \bibinfo
  {author} {\bibfnamefont {J.}~\bibnamefont {Echeverría}}, \bibinfo {author}
  {\bibfnamefont {E.}~\bibnamefont {Cremades}}, \bibinfo {author}
  {\bibfnamefont {F.}~\bibnamefont {Barragán}},\ and\ \bibinfo {author}
  {\bibfnamefont {S.}~\bibnamefont {Alvarez}},\ }\bibfield  {title} {\bibinfo
  {title} {Covalent radii revisited},\ }\href
  {https://doi.org/10.1039/b801115j} {\bibfield  {journal} {\bibinfo  {journal}
  {Dalton Transactions}\ ,\ \bibinfo {pages} {2832}} (\bibinfo {year}
  {2008})}\BibitemShut {NoStop}%
\bibitem [{\citenamefont {Yoshiasa}\ \emph {et~al.}(2003)\citenamefont
  {Yoshiasa}, \citenamefont {Murai}, \citenamefont {Ohtaka},\ and\
  \citenamefont {Katsura}}]{Yoshiasa2003}%
  \BibitemOpen
  \bibfield  {author} {\bibinfo {author} {\bibfnamefont {A.}~\bibnamefont
  {Yoshiasa}}, \bibinfo {author} {\bibfnamefont {Y.}~\bibnamefont {Murai}},
  \bibinfo {author} {\bibfnamefont {O.}~\bibnamefont {Ohtaka}},\ and\ \bibinfo
  {author} {\bibfnamefont {T.}~\bibnamefont {Katsura}},\ }\bibfield  {title}
  {\bibinfo {title} {Detailed structures of hexagonal diamond (lonsdaleite) and
  wurtzite-type bn},\ }\href {https://doi.org/10.1143/JJAP.42.1694} {\bibfield
  {journal} {\bibinfo  {journal} {Japanese Journal of Applied Physics}\
  }\textbf {\bibinfo {volume} {42}},\ \bibinfo {pages} {1694} (\bibinfo {year}
  {2003})}\BibitemShut {NoStop}%
\bibitem [{\citenamefont {Zhang}\ and\ \citenamefont
  {et~al.}(2024)}]{Zhang2024}%
  \BibitemOpen
  \bibfield  {author} {\bibinfo {author} {\bibfnamefont {Y.}~\bibnamefont
  {Zhang}}\ and\ \bibinfo {author} {\bibnamefont {et~al.}},\ }\bibfield
  {title} {\bibinfo {title} {New-generation ferroelectric alscn materials},\
  }\href {https://link.springer.com/article/10.1007/s40820-024-01441-1}
  {\bibfield  {journal} {\bibinfo  {journal} {Nano-Micro Letters}\ }\textbf
  {\bibinfo {volume} {16}},\ \bibinfo {pages} {227} (\bibinfo {year}
  {2024})}\BibitemShut {NoStop}%
\bibitem [{\citenamefont {Bhattarai}\ and\ \citenamefont
  {et~al.}(2024)}]{Bhattarai2024}%
  \BibitemOpen
  \bibfield  {author} {\bibinfo {author} {\bibfnamefont {B.}~\bibnamefont
  {Bhattarai}}\ and\ \bibinfo {author} {\bibnamefont {et~al.}},\ }\bibfield
  {title} {\bibinfo {title} {Effect of sc spatial distribution on the
  electronic and ferroelectric properties of alscn},\ }\href
  {https://doi.org/10.1039/D4MH00551A} {\bibfield  {journal} {\bibinfo
  {journal} {Materials Horizons}\ }\textbf {\bibinfo {volume} {11}},\ \bibinfo
  {pages} {5402} (\bibinfo {year} {2024})}\BibitemShut {NoStop}%
\bibitem [{\citenamefont {Resta}(1992)}]{Resta1992}%
  \BibitemOpen
  \bibfield  {author} {\bibinfo {author} {\bibfnamefont {R.}~\bibnamefont
  {Resta}},\ }\bibfield  {title} {\bibinfo {title} {Theory of the electric
  polarization in crystals},\ }\href
  {https://doi.org/10.1080/00150199208016065} {\bibfield  {journal} {\bibinfo
  {journal} {Ferroelectrics}\ }\textbf {\bibinfo {volume} {136}},\ \bibinfo
  {pages} {51} (\bibinfo {year} {1992})}\BibitemShut {NoStop}%
\bibitem [{\citenamefont {Vanderbilt}\ and\ \citenamefont
  {King-Smith}(1993)}]{Vanderbilt1993}%
  \BibitemOpen
  \bibfield  {author} {\bibinfo {author} {\bibfnamefont {D.}~\bibnamefont
  {Vanderbilt}}\ and\ \bibinfo {author} {\bibfnamefont {R.~D.}\ \bibnamefont
  {King-Smith}},\ }\bibfield  {title} {\bibinfo {title} {Electric polarization
  as a bulk quantity and its relation to surface charge},\ }\href
  {https://doi.org/10.1103/PhysRevB.48.4442} {\bibfield  {journal} {\bibinfo
  {journal} {Physical Review B}\ }\textbf {\bibinfo {volume} {48}},\ \bibinfo
  {pages} {4442} (\bibinfo {year} {1993})}\BibitemShut {NoStop}%
\bibitem [{\citenamefont {Resta}(1994)}]{Resta1994}%
  \BibitemOpen
  \bibfield  {author} {\bibinfo {author} {\bibfnamefont {R.}~\bibnamefont
  {Resta}},\ }\bibfield  {title} {\bibinfo {title} {Macroscopic polarization in
  crystalline dielectrics: the geometric phase approach},\ }\href
  {https://doi.org/10.1103/RevModPhys.66.899} {\bibfield  {journal} {\bibinfo
  {journal} {Reviews of Modern Physics}\ }\textbf {\bibinfo {volume} {66}},\
  \bibinfo {pages} {899} (\bibinfo {year} {1994})}\BibitemShut {NoStop}%
\bibitem [{\citenamefont {Dreyer}\ \emph {et~al.}(2016)\citenamefont {Dreyer},
  \citenamefont {Janotti}, \citenamefont {de~Walle},\ and\ \citenamefont
  {Vanderbilt}}]{Dreyer2016}%
  \BibitemOpen
  \bibfield  {author} {\bibinfo {author} {\bibfnamefont {C.~E.}\ \bibnamefont
  {Dreyer}}, \bibinfo {author} {\bibfnamefont {A.}~\bibnamefont {Janotti}},
  \bibinfo {author} {\bibfnamefont {C.~G.~V.}\ \bibnamefont {de~Walle}},\ and\
  \bibinfo {author} {\bibfnamefont {D.}~\bibnamefont {Vanderbilt}},\ }\bibfield
   {title} {\bibinfo {title} {Correct implementation of polarization constants
  in wurtzite materials and impact on iii-nitrides},\ }\href
  {https://doi.org/10.1103/PhysRevX.6.021038} {\bibfield  {journal} {\bibinfo
  {journal} {Physical Review X}\ }\textbf {\bibinfo {volume} {6}},\ \bibinfo
  {pages} {021038} (\bibinfo {year} {2016})}\BibitemShut {NoStop}%
\bibitem [{\citenamefont {Messi}\ \emph {et~al.}(2025)\citenamefont {Messi},
  \citenamefont {Patidar}, \citenamefont {Rodkey}, \citenamefont {Dräyer},
  \citenamefont {Trassin},\ and\ \citenamefont {Siol}}]{Messi2025}%
  \BibitemOpen
  \bibfield  {author} {\bibinfo {author} {\bibfnamefont {F.}~\bibnamefont
  {Messi}}, \bibinfo {author} {\bibfnamefont {J.}~\bibnamefont {Patidar}},
  \bibinfo {author} {\bibfnamefont {N.}~\bibnamefont {Rodkey}}, \bibinfo
  {author} {\bibfnamefont {C.~W.}\ \bibnamefont {Dräyer}}, \bibinfo {author}
  {\bibfnamefont {M.}~\bibnamefont {Trassin}},\ and\ \bibinfo {author}
  {\bibfnamefont {S.}~\bibnamefont {Siol}},\ }\bibfield  {title} {\bibinfo
  {title} {Ferroelectric alscn thin films with enhanced polarization and low
  leakage enabled by high-power impulse magnetron sputtering},\ }\href
  {https://doi.org/10.1063/5.0267904} {\bibfield  {journal} {\bibinfo
  {journal} {APL Materials}\ }\textbf {\bibinfo {volume} {13}},\ \bibinfo
  {pages} {051123} (\bibinfo {year} {2025})}\BibitemShut {NoStop}%
\bibitem [{\citenamefont {Drury}\ \emph {et~al.}(2022)\citenamefont {Drury},
  \citenamefont {Yazawa}, \citenamefont {Zakutayev}, \citenamefont {Hanrahan},\
  and\ \citenamefont {Brennecka}}]{Drury2022}%
  \BibitemOpen
  \bibfield  {author} {\bibinfo {author} {\bibfnamefont {D.}~\bibnamefont
  {Drury}}, \bibinfo {author} {\bibfnamefont {K.}~\bibnamefont {Yazawa}},
  \bibinfo {author} {\bibfnamefont {A.}~\bibnamefont {Zakutayev}}, \bibinfo
  {author} {\bibfnamefont {B.}~\bibnamefont {Hanrahan}},\ and\ \bibinfo
  {author} {\bibfnamefont {G.}~\bibnamefont {Brennecka}},\ }\bibfield  {title}
  {\bibinfo {title} {High-temperature ferroelectric behavior of al0.7sc0.3n},\
  }\href {https://doi.org/10.3390/mi13060887} {\bibfield  {journal} {\bibinfo
  {journal} {Micromachines}\ }\textbf {\bibinfo {volume} {13}},\ \bibinfo
  {pages} {887} (\bibinfo {year} {2022})}\BibitemShut {NoStop}%
\bibitem [{\citenamefont {Yasuoka}\ \emph {et~al.}(2020)\citenamefont
  {Yasuoka}, \citenamefont {Shimizu}, \citenamefont {Tateyama}, \citenamefont
  {Uehara}, \citenamefont {Yamada}, \citenamefont {Akiyama}, \citenamefont
  {Hiranaga}, \citenamefont {Cho},\ and\ \citenamefont
  {Funakubo}}]{Yasuoka2020}%
  \BibitemOpen
  \bibfield  {author} {\bibinfo {author} {\bibfnamefont {S.}~\bibnamefont
  {Yasuoka}}, \bibinfo {author} {\bibfnamefont {T.}~\bibnamefont {Shimizu}},
  \bibinfo {author} {\bibfnamefont {A.}~\bibnamefont {Tateyama}}, \bibinfo
  {author} {\bibfnamefont {M.}~\bibnamefont {Uehara}}, \bibinfo {author}
  {\bibfnamefont {H.}~\bibnamefont {Yamada}}, \bibinfo {author} {\bibfnamefont
  {M.}~\bibnamefont {Akiyama}}, \bibinfo {author} {\bibfnamefont
  {Y.}~\bibnamefont {Hiranaga}}, \bibinfo {author} {\bibfnamefont
  {Y.}~\bibnamefont {Cho}},\ and\ \bibinfo {author} {\bibfnamefont
  {H.}~\bibnamefont {Funakubo}},\ }\bibfield  {title} {\bibinfo {title}
  {Effects of deposition conditions on the ferroelectric properties of
  (al\textsubscript{1-x}sc\textsubscript{x})n thin films},\ }\href
  {https://doi.org/10.1063/5.0015281} {\bibfield  {journal} {\bibinfo
  {journal} {Journal of Applied Physics}\ }\textbf {\bibinfo {volume} {128}},\
  \bibinfo {pages} {114103} (\bibinfo {year} {2020})}\BibitemShut {NoStop}%
\bibitem [{\citenamefont {Song}\ \emph {et~al.}(2025)\citenamefont {Song},
  \citenamefont {Pradhan}, \citenamefont {Hu}, \citenamefont {Zhang},
  \citenamefont {Keneipp}, \citenamefont {Susner}, \citenamefont
  {Bhattacharya}, \citenamefont {Drndić}, \citenamefont {Olsson},\ and\
  \citenamefont {Jariwala}}]{Song2025}%
  \BibitemOpen
  \bibfield  {author} {\bibinfo {author} {\bibfnamefont {S.}~\bibnamefont
  {Song}}, \bibinfo {author} {\bibfnamefont {D.~K.}\ \bibnamefont {Pradhan}},
  \bibinfo {author} {\bibfnamefont {Z.}~\bibnamefont {Hu}}, \bibinfo {author}
  {\bibfnamefont {Y.}~\bibnamefont {Zhang}}, \bibinfo {author} {\bibfnamefont
  {R.~N.}\ \bibnamefont {Keneipp}}, \bibinfo {author} {\bibfnamefont {M.~A.}\
  \bibnamefont {Susner}}, \bibinfo {author} {\bibfnamefont {P.}~\bibnamefont
  {Bhattacharya}}, \bibinfo {author} {\bibfnamefont {M.}~\bibnamefont
  {Drndić}}, \bibinfo {author} {\bibfnamefont {R.~H.}\ \bibnamefont
  {Olsson}},\ and\ \bibinfo {author} {\bibfnamefont {D.}~\bibnamefont
  {Jariwala}},\ }\bibfield  {title} {\bibinfo {title} {Observation of giant
  remnant polarization in ultrathin alscn at cryogenic temperatures},\ }\href
  {http://arxiv.org/abs/2503.19491} {\bibfield  {journal} {\bibinfo  {journal}
  {ArXiv}\ } (\bibinfo {year} {2025})}\BibitemShut {NoStop}%
\bibitem [{\citenamefont {Jiang}\ \emph {et~al.}(2019)\citenamefont {Jiang},
  \citenamefont {Paillard}, \citenamefont {Vanderbilt}, \citenamefont {Xiang},\
  and\ \citenamefont {Bellaiche}}]{Jiang2019}%
  \BibitemOpen
  \bibfield  {author} {\bibinfo {author} {\bibfnamefont {Z.}~\bibnamefont
  {Jiang}}, \bibinfo {author} {\bibfnamefont {C.}~\bibnamefont {Paillard}},
  \bibinfo {author} {\bibfnamefont {D.}~\bibnamefont {Vanderbilt}}, \bibinfo
  {author} {\bibfnamefont {H.}~\bibnamefont {Xiang}},\ and\ \bibinfo {author}
  {\bibfnamefont {L.}~\bibnamefont {Bellaiche}},\ }\bibfield  {title} {\bibinfo
  {title} {Designing multifunctionality via assembling dissimilar materials:
  Epitaxial aln/scn superlattices},\ }\href
  {https://doi.org/10.1103/PhysRevLett.123.096801} {\bibfield  {journal}
  {\bibinfo  {journal} {Physical Review Letters}\ }\textbf {\bibinfo {volume}
  {123}},\ \bibinfo {pages} {096801} (\bibinfo {year} {2019})}\BibitemShut
  {NoStop}%
\bibitem [{\citenamefont {Lee}\ \emph {et~al.}(2025)\citenamefont {Lee},
  \citenamefont {Smaha}, \citenamefont {Brennecka}, \citenamefont {Haegel},
  \citenamefont {Gorai},\ and\ \citenamefont {Yazawa}}]{Lee2025algdn}%
  \BibitemOpen
  \bibfield  {author} {\bibinfo {author} {\bibfnamefont {C.-W.}\ \bibnamefont
  {Lee}}, \bibinfo {author} {\bibfnamefont {R.~W.}\ \bibnamefont {Smaha}},
  \bibinfo {author} {\bibfnamefont {G.~L.}\ \bibnamefont {Brennecka}}, \bibinfo
  {author} {\bibfnamefont {N.~M.}\ \bibnamefont {Haegel}}, \bibinfo {author}
  {\bibfnamefont {P.}~\bibnamefont {Gorai}},\ and\ \bibinfo {author}
  {\bibfnamefont {K.}~\bibnamefont {Yazawa}},\ }\bibfield  {title} {\bibinfo
  {title} {From prediction to experimental realization of ferroelectric
  wurtzite al\textsubscript{1-x}gd\textsubscript{x}n alloys},\ }\bibfield
  {journal} {\bibinfo  {journal} {APL Materials}\ }\textbf {\bibinfo {volume}
  {13}},\ \href {https://doi.org/10.1063/5.0251168} {10.1063/5.0251168}
  (\bibinfo {year} {2025})\BibitemShut {NoStop}%
\bibitem [{\citenamefont {Savant}\ \emph {et~al.}(2024)\citenamefont {Savant},
  \citenamefont {Gund}, \citenamefont {Nomoto}, \citenamefont {Maeda},
  \citenamefont {Jadhav}, \citenamefont {Lee}, \citenamefont {Ramesh},
  \citenamefont {Kim}, \citenamefont {Nguyen}, \citenamefont {Chen},
  \citenamefont {Casamento}, \citenamefont {Rana}, \citenamefont {Lal},
  \citenamefont {Xing},\ and\ \citenamefont {Jena}}]{Savant2024}%
  \BibitemOpen
  \bibfield  {author} {\bibinfo {author} {\bibfnamefont {C.}~\bibnamefont
  {Savant}}, \bibinfo {author} {\bibfnamefont {V.}~\bibnamefont {Gund}},
  \bibinfo {author} {\bibfnamefont {K.}~\bibnamefont {Nomoto}}, \bibinfo
  {author} {\bibfnamefont {T.}~\bibnamefont {Maeda}}, \bibinfo {author}
  {\bibfnamefont {S.}~\bibnamefont {Jadhav}}, \bibinfo {author} {\bibfnamefont
  {J.}~\bibnamefont {Lee}}, \bibinfo {author} {\bibfnamefont {M.}~\bibnamefont
  {Ramesh}}, \bibinfo {author} {\bibfnamefont {E.}~\bibnamefont {Kim}},
  \bibinfo {author} {\bibfnamefont {T.-S.}\ \bibnamefont {Nguyen}}, \bibinfo
  {author} {\bibfnamefont {Y.-H.}\ \bibnamefont {Chen}}, \bibinfo {author}
  {\bibfnamefont {J.}~\bibnamefont {Casamento}}, \bibinfo {author}
  {\bibfnamefont {F.}~\bibnamefont {Rana}}, \bibinfo {author} {\bibfnamefont
  {A.}~\bibnamefont {Lal}}, \bibinfo {author} {\bibfnamefont {H.~G.}\
  \bibnamefont {Xing}},\ and\ \bibinfo {author} {\bibfnamefont
  {D.}~\bibnamefont {Jena}},\ }\bibfield  {title} {\bibinfo {title}
  {Ferroelectric albn films by molecular beam epitaxy},\ }\href
  {https://doi.org/10.1063/5.0181217} {\bibfield  {journal} {\bibinfo
  {journal} {Applied Physics Letters}\ }\textbf {\bibinfo {volume} {125}},\
  \bibinfo {pages} {072902} (\bibinfo {year} {2024})}\BibitemShut {NoStop}%
\bibitem [{\citenamefont {Zhu}\ \emph {et~al.}(2022)\citenamefont {Zhu},
  \citenamefont {He}, \citenamefont {Hayden}, \citenamefont {Fan},
  \citenamefont {Yang}, \citenamefont {Maria},\ and\ \citenamefont
  {Trolier‐McKinstry}}]{Zhu2022}%
  \BibitemOpen
  \bibfield  {author} {\bibinfo {author} {\bibfnamefont {W.}~\bibnamefont
  {Zhu}}, \bibinfo {author} {\bibfnamefont {F.}~\bibnamefont {He}}, \bibinfo
  {author} {\bibfnamefont {J.}~\bibnamefont {Hayden}}, \bibinfo {author}
  {\bibfnamefont {Z.}~\bibnamefont {Fan}}, \bibinfo {author} {\bibfnamefont
  {J.~I.}\ \bibnamefont {Yang}}, \bibinfo {author} {\bibfnamefont
  {J.}~\bibnamefont {Maria}},\ and\ \bibinfo {author} {\bibfnamefont
  {S.}~\bibnamefont {Trolier‐McKinstry}},\ }\bibfield  {title} {\bibinfo
  {title} {Wake‐up in al\textsubscript{1-x}b\textsubscript{x}n ferroelectric
  films},\ }\href {https://doi.org/10.1002/aelm.202100931} {\bibfield
  {journal} {\bibinfo  {journal} {Advanced Electronic Materials}\ }\textbf
  {\bibinfo {volume} {8}},\ \bibinfo {pages} {2100931} (\bibinfo {year}
  {2022})}\BibitemShut {NoStop}%
\bibitem [{\citenamefont {Bellaiche}\ and\ \citenamefont
  {Zunger}(1998)}]{Bellaiche1998}%
  \BibitemOpen
  \bibfield  {author} {\bibinfo {author} {\bibfnamefont {L.}~\bibnamefont
  {Bellaiche}}\ and\ \bibinfo {author} {\bibfnamefont {A.}~\bibnamefont
  {Zunger}},\ }\bibfield  {title} {\bibinfo {title} {Effects of atomic
  short-range order on the electronic and optical properties of gaasn, gainn,
  and gainas alloys},\ }\href {https://doi.org/10.1103/PhysRevB.57.4425}
  {\bibfield  {journal} {\bibinfo  {journal} {Phys. Rev. B}\ }\textbf {\bibinfo
  {volume} {57}},\ \bibinfo {pages} {4425} (\bibinfo {year}
  {1998})}\BibitemShut {NoStop}%
\bibitem [{\citenamefont {George}\ \emph {et~al.}(2003)\citenamefont {George},
  \citenamefont {{\'I}{\~n}iguez},\ and\ \citenamefont
  {Bellaiche}}]{George2003}%
  \BibitemOpen
  \bibfield  {author} {\bibinfo {author} {\bibfnamefont {A.~M.}\ \bibnamefont
  {George}}, \bibinfo {author} {\bibfnamefont {J.}~\bibnamefont
  {{\'I}{\~n}iguez}},\ and\ \bibinfo {author} {\bibfnamefont {L.}~\bibnamefont
  {Bellaiche}},\ }\bibfield  {title} {\bibinfo {title} {Effects of atomic
  short-range order on the properties of perovskite alloys in their
  morphotropic phase boundary},\ }\href
  {https://doi.org/10.1103/PhysRevLett.91.045504} {\bibfield  {journal}
  {\bibinfo  {journal} {Phys. Rev. Lett.}\ }\textbf {\bibinfo {volume} {91}},\
  \bibinfo {pages} {045504} (\bibinfo {year} {2003})}\BibitemShut {NoStop}%
\bibitem [{\citenamefont {Prosandeev}\ and\ \citenamefont
  {Bellaiche}(2016)}]{Prosandeev2016}%
  \BibitemOpen
  \bibfield  {author} {\bibinfo {author} {\bibfnamefont {S.}~\bibnamefont
  {Prosandeev}}\ and\ \bibinfo {author} {\bibfnamefont {L.}~\bibnamefont
  {Bellaiche}},\ }\bibfield  {title} {\bibinfo {title} {Effects of atomic
  short-range order on properties of the
  pbmg\textsubscript{1/3}nb\textsubscript{2/3}o\textsubscript{3} relaxor
  ferroelectric},\ }\href {https://doi.org/10.1103/PhysRevB.94.180102}
  {\bibfield  {journal} {\bibinfo  {journal} {Physical Review B}\ }\textbf
  {\bibinfo {volume} {94}},\ \bibinfo {pages} {180102} (\bibinfo {year}
  {2016})}\BibitemShut {NoStop}%
\bibitem [{\citenamefont {Mourad}\ and\ \citenamefont
  {Czycholl}(2012)}]{mourad2012theory}%
  \BibitemOpen
  \bibfield  {author} {\bibinfo {author} {\bibfnamefont {D.}~\bibnamefont
  {Mourad}}\ and\ \bibinfo {author} {\bibfnamefont {G.}~\bibnamefont
  {Czycholl}},\ }\bibfield  {title} {\bibinfo {title} {Theory of band gap
  bowing of disordered substitutional ii--vi and iii--v semiconductor alloys},\
  }\href@noop {} {\bibfield  {journal} {\bibinfo  {journal} {The European
  Physical Journal B}\ }\textbf {\bibinfo {volume} {85}},\ \bibinfo {pages}
  {153} (\bibinfo {year} {2012})}\BibitemShut {NoStop}%
\bibitem [{\citenamefont {Roy}\ \emph {et~al.}(2024)\citenamefont {Roy},
  \citenamefont {Sieradzki}, \citenamefont {Rondinelli},\ and\ \citenamefont
  {McCue}}]{Roy2024}%
  \BibitemOpen
  \bibfield  {author} {\bibinfo {author} {\bibfnamefont {A.}~\bibnamefont
  {Roy}}, \bibinfo {author} {\bibfnamefont {K.}~\bibnamefont {Sieradzki}},
  \bibinfo {author} {\bibfnamefont {J.~M.}\ \bibnamefont {Rondinelli}},\ and\
  \bibinfo {author} {\bibfnamefont {I.~D.}\ \bibnamefont {McCue}},\ }\bibfield
  {title} {\bibinfo {title} {Effect of chemical short-range order and
  percolation on passivation in binary alloys},\ }\href
  {https://doi.org/10.1103/PhysRevB.110.085420} {\bibfield  {journal} {\bibinfo
   {journal} {Physical Review B}\ }\textbf {\bibinfo {volume} {110}},\ \bibinfo
  {pages} {085420} (\bibinfo {year} {2024})}\BibitemShut {NoStop}%
\end{thebibliography}%

\end{document}